\documentclass[a4paper,10pt]{article}
\usepackage[cm]{fullpage}
\usepackage{authblk}
\usepackage{amsmath,longtable,lscape,natbib}   % txfonts
\usepackage[font=small,labelfont=bf,width=160mm]{caption}
\usepackage[pdftex]{graphicx}
\usepackage{subfigure}
\usepackage{verbatim}
\usepackage{url}
\usepackage{bm}
\usepackage{bigints}
\usepackage{paralist}

\usepackage{amssymb}  % Mathesymbole
\begin{document}
\setlength{\skip\footins}{5mm}
%\begin{comment}

%%%%%%%%%%%%%%%%%%%%%%%%%%%%%%%%%%%%%%%%%
\title{\bf Theoretical study of ionization profiles of molecular clouds near supernova remnants \\ \vspace{10pt}
\large{Tracing the hadronic origin of GeV gamma radiation}}
% \subtitle{\textbf{Tracing the hadronic origin of GeV gamma radiation}}

\author[1]{Florian Schuppan\footnote{Corresponding author. Contact: florian.schuppan@rub.de, phone: +49-234-3222329}$^,$}%\,\,, 
\author[2]{Christian R\"oken}%, and 
\author[1]{Julia Becker Tjus}% \\
\affil[1]{\noindent \small{Ruhr-Universit\"at Bochum, Fakult\"at f\"ur Physik \& Astronomie, Institut f\"ur Theoretische Physik IV, \newline 
44780 Bochum, Germany}\vspace{0.1cm}} %\\
\affil[2]{\noindent \small{Universit\"at Regensburg, Fakult\"at f\"ur Mathematik, 93040 Regensburg, Germany}} 
\renewcommand\Authands{, and }
 \date{July 9, 2014}
\maketitle

\abstract
   \noindent {\textit{Context}: Over the past few years, signatures of supernova remnants have been detected in gamma rays, particularly those that are known to be 
    associated with molecular clouds. Whether these gamma rays are generated by cosmic-ray electrons 
    emitting bremsstrahlung or experiencing inverse Compton scattering, or by cosmic-ray protons interacting 
    with ambient hydrogen is usually not known. The detection of hadronic ionization signatures 
    in spatial coincidence with gamma-ray signatures can help to unambiguously identify 
    supernova remnants as sources of cosmic-ray protons. \\
   \textit{Aims}: Our central aim is to develop a method to investigate whether the gamma rays are formed by cosmic-ray protons. 
   To achieve this goal, we derived the position-dependent cosmic-ray-induced and photoinduced ionization rates. \\
   \textit{Methods}: To calculate hadronic signatures from cosmic-ray-induced ionization to examine 
    the origin of the observed gamma rays from molecular clouds associated with supernova remnants, 
    we solved analytically 
    the transport equation for cosmic-ray protons propagating in a molecular cloud, including the 
    relevant momentum-loss processes, and determined the proton flux at any penetration depth into 
    the cloud. \\ 
   \textit{Results}: Because the solution of the transport equation is obtained for arbitrary source functions, it can be used for a 
   variety of supernova remnants. We derived the corresponding theoretical ionization rate as a function of the penetration depth 
   from the position-dependent proton flux, and compared it with photoinduced ionization profiles for available 
   X-ray data in a case study with four supernova remnants associated with molecular clouds. Three of the remnants show a 
   clear dominance of the hadronically induced ionization rate, while for one remnant, X-ray emission seems to 
   dominate by a factor of 10. \\ 
   \textit{Conclusions}: This is the first derivation of position-dependent profiles for cosmic-ray-induced ionization with an analytic solution for 
   arbitrary cosmic-ray source spectra. The cosmic-ray-induced ionization has to be compared with photoionization for 
   strong X-ray sources. For this purpose, measurements of X-ray spectra from supernova remnant shocks in the sub-keV to keV 
   domain are necessary for a proper comparison. For sources dominated by cosmic-ray-induced ionization (e.g., W49B), the 
   ionization profiles can be used in the future to map the spatial structure of hadronic gamma rays and 
   rotation-vibrational lines induced by cosmic-ray protons. With instruments such as ALMA for the line signatures 
   and CTA for the gamma-ray detection, this correlation study will help to identify sources of hadronic cosmic rays. }

% \authorrunning{F.\ Schuppan et al.}
% \titlerunning{Theoretical study of ionization profiles}   

%================================================
\section{Introduction\label{intro}}
%================================================
Supernova remnants (SNRs) are the main candidates for accelerating Galactic cosmic rays (CRs) \citep[see, e.g.,][]{baade1934,ackermann2013}. 
Particularly interesting are SNRs associated with molecular clouds (MCs) that are illuminated by CR protons 
accelerated at the SNR shock front, leading to the emission of gamma rays. Currently, there 
are about ten of these objects known to emit photons at GeV energies with typical 
values of $100\,\mathrm{MeV}\lesssim\,E_\gamma\,\lesssim\,10\,\mathrm{TeV}$ detected with FermiLAT  
\citep[see, e.g.,][]{abdo(W51C)2009, abdo(W44)2010} and H.E.S.S.\ \citep[see, e.g.,][]{aharonian2008}. 
Generally, there are three processes that might lead to the production of these gamma rays: 
Bremsstrahlung or inverse Compton scattering from CR electrons, or the decay of neutral pions that are formed 
via inelastic scattering of CR protons on ambient hydrogen. If only the 
detected gamma-ray spectrum is available, the processes causing the gamma radiation cannot be determined definitely 
from these observations since they neither allow distinguishing a leptonic from a hadronic (proton) scenario, 
nor do they help telling apart bremsstrahlung from inverse Compton scattering. %\\

Although sometimes estimates for the magnetic field strength or the seed photon field for 
inverse Compton scattering, or deductions of average hydrogen densities from observations can be made, 
there is still plenty of room for interpretation regarding the origin of the gamma radiation. 
One way to unambiguously recognize pion production as the source of the gamma rays is the detection of 
neutrinos formed by the decay of charged pions, which are generated via inelastic scattering of 
CR protons on ambient hydrogen at a fixed ratio to the formation of neutral pions \citep{particledatabook2010}. 
A first evidence for astrophysical neutrinos was recently announced by the IceCube collaboration \citep{icecube2013}. 
But at this point, the observed flux of extraterrestrial neutrinos is diffuse, and the statistics does not allow 
statements on possible point sources. 
Another way to indirectly explore whether the origin of the detected gamma radiation is of hadronic nature is given 
by the idea that if the gamma rays are induced by high-energy CR protons ($E_{\mathrm{p}}\gtrsim1\,$GeV), 
one also expects CR protons of lower energy, which, because of the threshold energy for pion production, do not 
contribute to the gamma flux \citep{becker2011}. 
These low-energy protons are quite efficient in ionizing the MCs, even more efficient than CR electrons \citep{padovani2009}, 
because of two effects. First, the ionization cross-section for electrons is lower than the one for protons. 
Second, the lower rest mass of the electrons causes them (a) to be deflected by electromagnetic fields to a 
higher degree, and (b) to lose their energy more rapidly than CR protons in interactions with matter. 
Therefore, the CR electrons cannot penetrate an MC deep enough to contribute significantly to the overall 
ionization rate in its interior. 
Hence, among all the scenarios proposed to explain the observed gamma radiation, only the hadronic one 
would imply effective ionization of MCs, particularly in their inner regions. A 
correlation study of both high-energy gamma rays and low-energy ionization signatures in spatial coincidence 
might therefore provide a strong support for a hadronic scenario for the generation of gamma rays. 

To be able to attribute ionization signatures to low-energy CR protons, other possible ionization 
sources need to be ruled out, in particular UV photons and X-rays. UV photons are typically absorbed quite effectively already at 
low column densities of traversed matter, therefore, they play a crucial role in the total ionization 
only at the outer regions of the clouds \citep{tielens2010}. X-rays, however, can penetrate the clouds 
considerably deeper and are also quite effective in their ionization. Consequently, photoionization 
is the process inside MCs that is required to be exceeded by CR-induced ionization to enable one to attribute 
detectable ionization signatures to CRs (here, the term \textit{CR-induced ionization} 
denotes ionization induced only by primary CR protons and the corresponding secondary electrons).  
These signatures can be distinguished in terms of their variation in the column-density dependence in the cloud. 
In this study, the ionization rates induced by both processes are derived as functions of the penetration depth 
into the cloud, thus providing a crucial clue to the dominant processes that contribute to the 
formation of gamma rays from SNRs associated with MCs. 
\\
\\
The paper is organized as follows: first, the relation between gamma-ray observations and a potential underlying 
CR proton spectrum is described in Sect.\ \ref{CR_prot}. In Sect.\ \ref{TE}, the transport equation for the propagation 
of the CR protons into and inside the MCs is introduced, and both the general analytic solution and a solution for specific 
choices of the diffusion coefficient, momentum losses, and source function are given. 
For four specific SNR-MC systems, the position-dependent ionization rates induced by CR protons are derived, using the 
solution of the transport equation, in Sect.\ \ref{CR_ion_prof}. 
Section \ref{X-ray} deals with the photon fluxes both at the surfaces and inside the clouds of these systems and the corresponding 
position-dependent photoionization rates. 
The ionization rates induced by photons and CR protons are compared in Sect.\ \ref{results}. 
Finally, in Sect.\ \ref{conclusions}, we give a brief overview of how ionization rates affect MCs and how they 
can be detected are provided. We also provide conclusions from the results from Sect.\ \ref{results} and 
give an outlook for future work. 

%\end{comment}

%================================================
\section{Cosmic ray proton spectrum\label{CR_prot}}
%================================================
The most established processes for the acceleration of CRs are Fermi acceleration \citep{fermi1949} of first and 
second order, in particular diffusive shock acceleration \citep[see, e.g.,][]{axford1977, bell1978b, 
bell1978, jokipii1987, schlickeiser1989a, schlickeiser1989b}. Within 
specific parameter domains, these can account for typical CR proton spectra from isolated sources. Therefore, 
it is not possible to determine which acceleration process is actually at work at a certain source. Usually, this 
poses a profound problem for the construction of theoretical models for specific objects. Nevertheless, it can be circumvented when the CR 
proton spectrum directly at the interaction region is derived from observational data. %\\

In this work, a hadronic scenario for the generation of gamma rays in the SNR-MC systems 
is considered, and, using observed gamma-ray spectra, the proton fluxes at the gamma-ray emission regions 
are constructed assuming that the detected gamma radiation is caused by the decay of neutral pions formed 
in proton-proton interactions of CR protons with ambient hydrogen 
and that homogeneous volumes enclosed by the SNR shock fronts are 
the source regions of isotropic CR proton emission and that the MC surfaces are in contact with the shock 
fronts. 
For a parametric description of these proton fluxes, 
their parameters are chosen in such a way that the resulting gamma-ray emission is compatible 
with the observations. Models for the interaction of CR protons with ambient matter are discussed in \cite{kelner2006} 
and \cite{kamae2006}, leading to parametric descriptions of the CR proton injection fluxes into the cloud 
with constrained kinetic proton energies $E_{\mathrm{p}}\gtrsim280\,$MeV. This lower boundary comes from 
the minimal energy required for the formation of neutral pions, which will decay into the detected 
gamma rays for a hydrogen target at rest. However, since only the low-energy CR protons are able 
to efficiently ionize the MCs, the fluxes are extrapolated toward lower energies 
under the assumption that they follow the same power-law behavior as in the energy ranges described by observations. 
The a priori unknown shape of the CR proton spectrum at low energies is a major problem, but the extrapolation performed 
here is appropriate for several reasons. First, it spans less than one order of magnitude in terms of the 
particle momentum and, therefore, it is close to the momentum domain covered by observational data. 
Second, there is no apparent change in the observed spectral shape of the CR protons toward lower energies. 
Third, observations of ionization signatures from \cite{indriolo2009} indicate that a CR proton 
flux that increases toward lower proton energies is required in order to explain these detections. Thus, the uncertainty associated 
with the extrapolation is reasonably small compared to other uncertainties in studies of SNRs near MCs, such as the 
average hydrogen density of the cloud or the volume-filling factor of the volume enclosed by the SNR shock front. 
The propagation of the CR protons inside the MCs is modeled in terms of a transport equation containing a general injection 
spectrum, momentum-dependent scalar diffusion, and ionization, Coulomb and adiabatic deceleration losses. 
The MCs are of homogeneous hydrogen density and of arbitrary shape. 

%\newpage

%================================================
\section{Transport equation \label{TE}}
%================================================
In this section, the relevant transport equation for the CR protons is presented, the 
components are explained and discussed, and the solution is given in analytic form. 
The transport equation describing the temporal evolution of primary CR protons inside an MC located 
in the vicinity of an SNR reads
\begin{equation}
\frac{\partial n_{\mathrm{p}}(\vec{r},p,t)}{\partial t} - D(p) \Delta n_{\mathrm{p}}(\vec{r},p,t) 
- \frac{\partial}{\partial p}\Big(b(p) \cdot n_{\mathrm{p}}(\vec{r},p,t) \Big) = Q(\vec{r},p,t),
\label{transp_eq_large}
\end{equation}
where $n_{\mathrm{p}}(\vec{r},p,t)$ is the differential number density of protons at the point $\vec{r}$ with momentum $p$ at time $t$, 
$D(p)$ is the scalar momentum-dependent diffusion coefficient, $\Delta$ is the Laplace operator, 
$b(p)$ is the momentum-loss rate, and $Q(\vec{r},p,t)$ is a general source function for a supernova shock front 
that accelerates CR protons. 
Note that this equation does not hold for highly relativistic momenta, since in that case the diffusion coefficient 
can no longer be described by a scalar quantity. Furthermore, convection and advection of particles as well 
as diffusion in momentum space are not considered. 
The Green's function of this transport equation is calculated in detail in Appendix A using 
Laplace transformations and Duhamel's principle \citep{duhamel1838, duhamel}, 
yielding 
\begin{equation}
G(\vec{r},p,t\,|\,\vec{r}_0,p_0) = \frac{\Theta(p_0 - p) \delta\left(t + \int_{p_0}^p{b(p')^{-1}\,{\mathrm d}p'}\right)}
{b(p) \cdot \left(4 \pi \int_{p}^{p_0}{D(p')/b(p')\,{\mathrm d}p'}\right)^{3/2}} 
\exp \left(\frac{-(\vec{r} - \vec{r}_0)^2}{4 \int_{p}^{p_0}{D(p')/b(p')\,{\mathrm d}p'}} \right),
\label{eq:G_sol}
\end{equation}
where $\Theta(\cdot)$ is the Heaviside step function and $\delta(\cdot)$ is the Dirac distribution. 
The general solution $n_{\mathrm{p}}(\vec{r},p,t)$ 
for an arbitrary source function $Q(\vec{r}_0,p_0,t_0)$ can be obtained by convolving 
the fundamental solution (\ref{eq:G_sol}) with the source function 
\begin{equation}
n_{\mathrm{p}}(\vec{r},p,t) = \iiint{G(\vec{r},p,t\,|\,\vec{r}_0,p_0)Q(\vec{r}_0,p_0,t_0)\,{\mathrm d}t_0\,{\mathrm d}^3r_0\,{\mathrm d}p_0}.
\label{convol_G_Q}
\end{equation}
This differential CR proton number density can also be applied to many other astrophysical situations with scalar, momentum-dependent 
diffusion and any type of momentum losses, such as stellar winds, CR diffusion in the interstellar medium, or gamma-ray bursts.

Two specific momentum-loss processes in SNR-MC systems are considered: Coulomb 
collision losses (i.e., ionization or excitation of interstellar matter) and adiabatic deceleration (i.e., 
attenuation of the CR number density due to the expansion of the shock front with a momentum-loss timescale independent 
of the CR particle momentum). 
The loss rate for Coulomb collisions \citep{lerche-schlick1982} can be given in the approximated form 
\begin{equation}
b_{\mathrm{cc}}(p) = 5 \cdot 10^{-19} Z^2 \bigg(\frac{n_{\mathrm{H}}}{{\mathrm{cm}^{-3}}} \bigg) 
\bigg(\frac{p}{Mc} \bigg)^{-2} \Bigg(11.3 + 2 
\ln \bigg(\frac{p}{Mc} \bigg) \Bigg)~\mathrm{eV~cm^{-1}},
\label{eq:Coul_large}
\end{equation}
where $Z$ is the charge number of the incoming CR particle, $n_{\mathrm{H}}$ is the hydrogen density 
of the medium, $M$ is the particle rest mass, and $c$ denotes the speed of light. 
Here, only protons (i.e., $Z=1$ and $M=m_{\mathrm p}$) of a minimum momentum of $0.15\,m_{\mathrm p}c$, corresponding to a 
kinetic energy of 10~MeV, and a maximum momentum of $0.86\,m_{\mathrm p}c$, corresponding to a kinetic 
energy of 280~MeV, are considered. 
Protons of lower momentum can be neglected since they do not penetrate the cloud deep enough in order to provide an 
observable contribution to the ionization rate. Protons of higher momentum can be neglected on the one hand, 
because of the rapid decrease of the ionization cross-section with increasing momentum 
\citep{rudd1985} and, on the other hand, because they can form pions by inelastic proton-proton scattering 
(the momentum of $\sim0.86\,m_{\mathrm p}c$ corresponds to the threshold kinetic energy 
required for protons to produce pions by proton-proton interactions), which renders them unavailable 
for the ionization of the MC. 
Imposing these limits on the CR proton momentum, the logarithmic contribution in 
\mbox{Eq.\ (\ref{eq:Coul_large})} is small compared to the constant and is therefore, as an approximation, 
disregarded. Then, the loss rate for Coulomb collisions simplifies to
\begin{equation}
b_{\mathrm{cc}}(p) = a_{\mathrm{cc}} \left(\frac{p}{m_{\mathrm p}c} \right)^{-2}
\label{eq:Coul}
\end{equation}
with
\begin{equation}
a_{\mathrm{cc}} = 5.65 \cdot 10^{-18} \left(\frac{n_{\mathrm{H}}}{{\mathrm{cm}^{-3}}} \right)~\mathrm{eV~cm^{-1}}.
\label{a_cou}
\end{equation} 
The loss rate for adiabatic deceleration at a shock velocity $\vec{V}$ can be expressed as 
\begin{equation}
b_{\mathrm{ad}}(p) = \frac{1}{3}\,p~\nabla\cdot\vec{V}.
\label{eq:adiab_large}
\end{equation}
Assuming a constant, spherical expansion of the shock front, the shock velocity reads
\begin{equation}
\vec{V} = V_0 \, \vec{e}_R\,,
\end{equation}
where $V_0$ is a constant, and $R$ is the shock radius. Hence, the divergence of the shock 
velocity becomes 
\begin{equation}
\nabla\cdot\vec{V} = \frac{1}{R^2} {\partial}_R \big(R^2 V_0 \big) = 2 \frac{V_0}{R},
\label{eq:div_V}
\end{equation}
and the loss rate for adiabatic deceleration (\ref{eq:adiab_large}) yields 
\begin{equation}
b_{\mathrm{ad}}(p) = a_{\mathrm{ad}} \left(\frac{p}{m_{\mathrm p}c} \right)
\label{eq:adiab_short}
\end{equation}
with
\begin{equation}
a_{\mathrm{ad}} = 2.085 \cdot 10^{-2} \left(\frac{V_0/R}{{\mathrm s^{-1}}}\right)~{\mathrm{ eV~cm^{-1}}}.
\label{a_ad}
\end{equation}
Note that for a constant shock velocity the quantity $V_0/R$ can be identified with the reciprocal age of 
the SNR, $t_{\mathrm{age}}$, thus, providing a reasonable estimate of the divergence 
of the shock velocity, which is otherwise extremely difficult to obtain from observations. \\
\begin{figure}[h]
\centering
 \includegraphics[width=0.5\columnwidth]{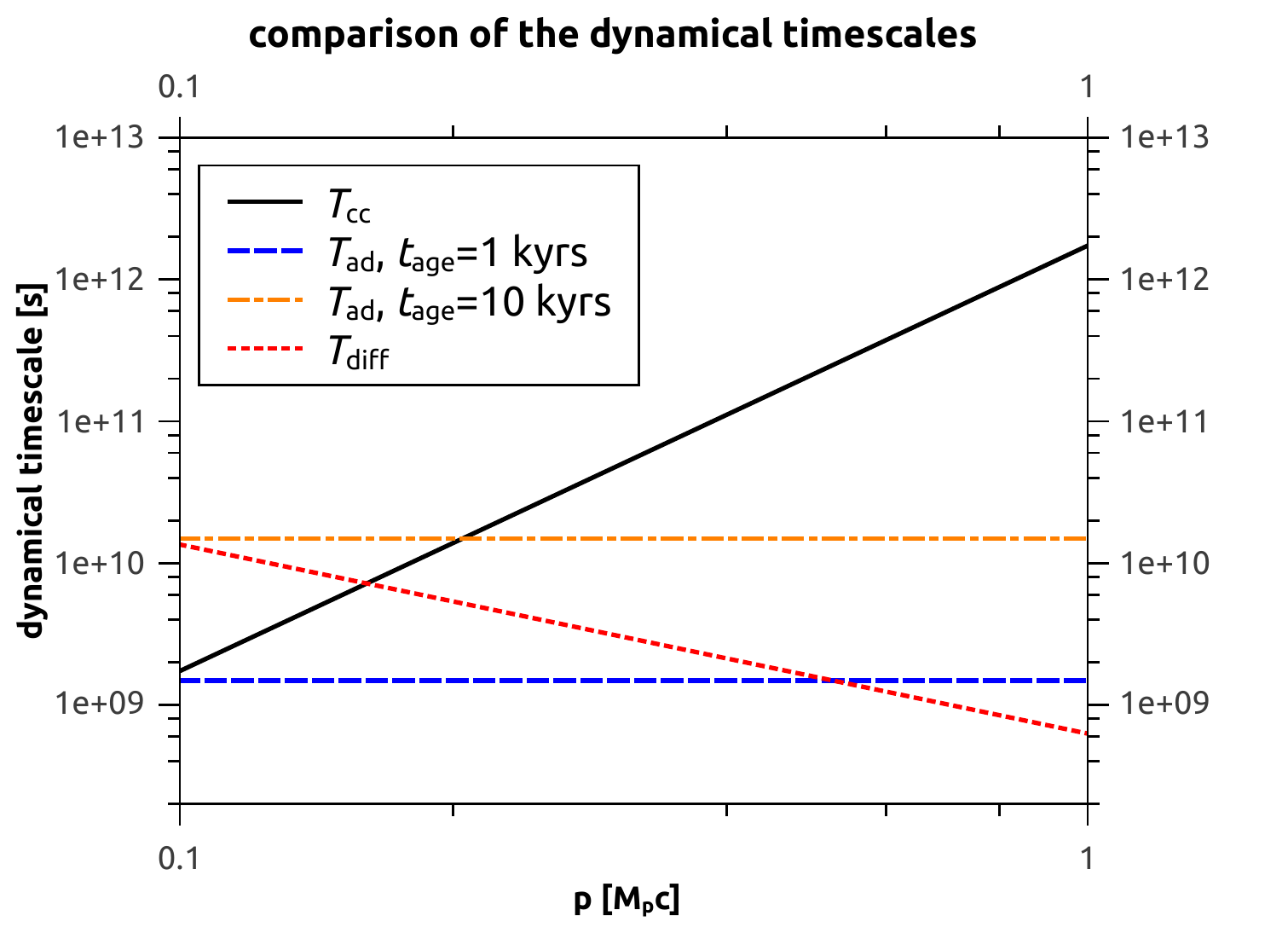}
\caption{Comparison of the dynamical timescales for Coulomb losses, adiabatic deceleration, and diffusion
 with the typical values $n_{\mathrm{H}}=100\,\mathrm{cm^{-3}}$, $t_{\mathrm{age}}=10\,\mathrm{kyr}$, and 
 a diffusion length $l=2\,\mathrm{pc}$. \label{timescales}}
\end{figure} 

\noindent In Fig.\ \ref{timescales}, the dynamical timescales for both Coulomb losses (solid, black line) and 
adiabatic deceleration (long-dashed, blue line for a young SNR and long-short-dashed, orange line for a middle-aged SNR), 
as well as the diffusion timescale (short-dashed, red line), as functions of the particle momentum $p$, are depicted. 
The various timescales are given by $T_{\mathrm{cc}}=(\mathrm{d}p/\mathrm{d}t)^{-1}_{\mathrm{cc}}p$, 
$T_{\mathrm{ad}}=(\mathrm{d}p/\mathrm{d}t)^{-1}_{\mathrm{ad}}p$, and $T_{\mathrm{diff}}=l^2/(2D(p))$, 
where $l$ is the diffusion length, and the diffusion coefficient for the SNR-MC systems is 
$D(p)=10^{28}\cdot(p/(m_{\mathrm{p}}c))^{4/3}\,\mathrm{cm^2\,s^{-1}}$ with a power-law index of $4/3$. 
Despite the specific choice of this power-law index, any power-law dependence of the diffusion coefficient 
on the particle momentum allows for analytic evaluation of the solution of the transport equation with the 
momentum losses considered here. 
The general form of the diffusion coefficient for a Kolmogorov spectrum in a statistically isotropic random 
magnetic field \citep[see, e.g.,][]{ptuskin2006} is $D(p)\propto v\,p^{1/3}$, where $v$ is the CR particle speed. 
In the non-relativistic limit, $v \propto p$ and, therefore, $D(p)\propto p^{4/3}$, yielding a reasonable 
approximation of the diffusion coefficient for low momenta. Since the kinetic energies of the CR protons considered in 
this work also exceed values that justify the non-relativistic treatment, 
the assumption $D(p)\propto p^{4/3}$ leads to an overestimate of the diffusion effect for particles with momenta 
exceeding the non-relativistic regime. This 
results in an overestimated ionization rate at large penetration depths, because the time a CR particle takes 
to reach a certain penetration depth decreases with an increasing diffusion coefficient, while at small penetration depths, 
the resulting ionization rate is underestimated. 
Apart from Kolmogorov spectra, Iroshnikov-Kraichnan spectra can be applied as well, by adapting 
the power-law index of the diffusion coefficient. When Goldreich-Sridhar spectra provide the adequate description, 
the diffusion coefficient splits into a parallel and perpendicular component relative to the mean magnetic field 
orientation.
The values chosen for the average hydrogen densities of the MCs and for the diffusion coefficient in Fig.\ \ref{timescales} are 
the same as those used in the data analysis in Sect.\ \ref{CR_ion_prof}. For 
middle-aged SNRs, losses from Coulomb collisions for $p \lesssim0.2\,m_{\mathrm{p}}c$ occur on shorter timescales than 
adiabatic deceleration losses. Thus, they dominate momentum losses of the CR protons in this regime, while at 
higher momenta, the losses from adiabatic deceleration prevail. For young SNRs, this transition is shifted toward 
lower momenta of $p\approx 0.1\,m_{\mathrm{p}}c$, because the dynamical timescale for adiabatic deceleration 
is directly proportional to the age of an SNR, and, therefore, the momentum loss from adiabatic deceleration 
for those SNRs dominates for all momenta covered by this study. 
The diffusion timescale indicates how long a particle with given momentum $p$ takes to diffuse $2\,\mathrm{pc}$, 
ignoring the effect of momentum losses. Figure \ref{timescales} indicates that for young objects, particles with a momentum 
$p\lesssim0.5\,m_{\mathrm{p}}c$ lose all their kinetic energy due to adiabatic deceleration losses before reaching a 
penetration depth of 2\,pc, while for middle-aged SNRs, this only happens for low-momentum particles 
with $p\lesssim0.15\,m_{\mathrm{p}}c$. The maximum penetration depth of the latter is governed by Coulomb losses. 
Using both Eqs.\ (\ref{a_cou}) and (\ref{a_ad}), the total momentum-loss rate can be expressed as 
\begin{equation}
b(p) = a_{\mathrm{ad}} \left(\frac{p}{m_{\mathrm p}c} \right) 
+ a_{\mathrm{cc}} \left(\frac{p}{m_{\mathrm p}c} \right)^{-2}.
\label{eq:loss_rate}
\end{equation}
Moreover, the source function $Q(\vec{r}_0,p_0,t_0)$ is constructed such that it reflects the geometry of 
the SNR-MC system and the energy dependence of the incident CR proton flux. This flux is derived from observations 
for four specific SNRs associated with MCs that show gamma-ray emission and for which data samples from spectral measurements 
in the X-ray energy range exist. Under the assumptions that the CR protons are constantly and isotropically emitted from 
the homogeneous region enclosed by the SNR shock front and that the MC is located directly 
adjacent to the emission volume, the source function is modeled in the specific form 
\begin{align}
 Q(\vec{r}_0,p_0,t_0) &= Q_{\mathrm{norm}}\cdot Q_{p}(p_0)\cdot\big[\Theta\left(x_0+l_{\mathrm{c}}/2\right)-\Theta\left(x_0-l_{\mathrm{c}}/2\right)\big] %\nonumber \\
 \cdot\big[\Theta\left(y_0+l_{\mathrm{c}}/2\right)-\Theta\left(y_0-l_{\mathrm{c}}/2\right)\big] \nonumber \\
 &\cdot\big[\Theta(z_0+l_{\mathrm{c}})-\Theta(z_0)\big]\cdot\big[\Theta(t_0)-\Theta(t_0-1)\big],
\label{source}
\end{align}
where $Q_{\mathrm{norm}}$ denotes a normalization constant and $Q_{p}(p_0)$ is the spectral shape of the 
low-energy CR protons in terms of the particle momentum $p_0$, describing emission that is constant over a time interval of 
one second from a cubic emission volume with edge length $l_{\rm c}$, seen by an observer located at the center of the face 
of the emission volume that is the boundary surface of the SNR shock front and the MC, with a coordinate system such that the 
positive Cartesian $z$-axis is normal to this face and points into the cloud. The volume of the emission region, $l_{\mathrm{c}}^3$, is 
modeled as cubic for numerical feasibility and adapted to the spherical emission volume used in the modeling process 
of the gamma rays in Sect.\ \ref{CR_ion_prof}. 

Performing the convolution in Eq.\ (\ref{convol_G_Q}) for the source function (\ref{source}) yields the 
differential CR proton number density at all positions $\vec{r}$ inside the MC with $z\ge0$ at any time $t\ge0$ and 
for all particle momenta $p\in[0.15\,m_{\mathrm{p}}c,\, 0.86\,m_{\mathrm{p}}c] \le p_0$ 
\begin{align} \label{eq:sol_transp_eq}
n_{\mathrm{p}}(\vec{r},p,t) &= \frac{Q_{\mathrm{norm}}\cdot\left(a_{\mathrm{cc}}+a_{\mathrm{ad}}
 \big(\mathfrak{p}_0^{\mathrm{zero}}(p,t)\big)^3\right)\cdot Q_{p}\big(\mathfrak{p}_0^{\mathrm{zero}}(p,t)\big)}
 {8\cdot b(p)  \cdot \big(\mathfrak{p}_0^{\mathrm{zero}}(p,t)\big)^2} \\
&\cdot \prod_{h=x,y,z+l_{\mathrm{c}}/2}\left[\sum_{j=0}^{1}\mathrm{erf}\left(\frac{\left(l_{\mathrm{c}}/2 + (-1)^j\cdot h\right)\sqrt{a_{\mathrm{cc}}(3+k)}}
 {\sqrt{4 D_0 \, m_{\mathrm p}c %\, 
 \left[\left(\mathfrak{p}' \right)^{3+k} \, _{2}F_{1}\left(1, 1+\frac{k}{3}; 2 + \frac{k}{3}; -a 
 \cdot \left(\mathfrak{p}' \right)^3 \right)\right]_{\mathfrak{p}' = p/(m_{\mathrm{p}}c)}^{\mathfrak{p}_0^{\mathrm{zero}}(p,t)}}} \right) \right], \nonumber
\end{align}
where $a = a_{\mathrm{ad}} / a_{\mathrm{cc}}$, $k$ is the power-law index of the diffusion coefficient (for the data 
analysis, $k=4/3$), erf$(\cdot)$ and $_{2}F_{1}(\cdot\,,\cdot\,;\cdot\,;\cdot)$ denote the error function 
and the hypergeometric function, respectively, and 
$$\mathfrak{p}_0^{\mathrm{zero}}(p,t) = \Bigg(\left(\frac{p}{m_{\mathrm{p}}c}\right)^3\cdot\exp\left(\frac{3a_{\mathrm{ad}}t}{m_{\mathrm p}c}\right) 
 + \frac{a_{\mathrm{cc}}}{a_{\mathrm{ad}}}\left(\exp\left(\frac{3a_{\mathrm{ad}}t}{m_{\mathrm p}c}\right)-1\right)\Bigg)^{1/3}.$$ 
For details, see Appendix B. 
In a next step, the undetermined normalization constant $Q_{\mathrm{norm}}$ and the momentum-dependent spectral function $Q_p$ 
are modeled such that they are adapted to the CR proton fluxes, derived from gamma-ray observations, in the SNR-MC systems outside the 
clouds, yielding initial boundary values of the differential CR proton number density at the cloud surfaces. 
Then, Eq.\ (\ref{eq:sol_transp_eq}) can be used to determine the CR proton fluxes inside the MCs that are 
necessary for calculating the ionization rates.

%%%%%%%%%%%%%%%%%%%%%%%%%%%%%%%%%%%%%%%%%%%%%%%%%%%%
\section{CR-induced ionization profiles}\label{CR_ion_prof}
%%%%%%%%%%%%%%%%%%%%%%%%%%%%%%%%%%%%%%%%%%%%%%%%%%%%
The CR-induced ionization rate of molecular hydrogen in MCs is computed for four specific objects: W49B, W44, 3C~391, 
and CTB~37A, following \citep{padovani2009}, 
\begin{equation}
  \zeta^{{\mathrm H}_2}(\vec{r},t) = \left(1 +\phi\right)\int_{E_{\min}}^{E_{\max}} 
 \frac{{\mathrm d}^3N_{\mathrm p}(\vec{r},E_{\mathrm{p}},t)}{\mathrm{d}E_{\mathrm{p}}\,{\mathrm d}A\,{\mathrm d}t}\; 
 \sigma_{{\mathrm{ion}}}^{{\mathrm H}_2}(E_{\mathrm{p}}) \,{\mathrm d}E_{\mathrm{p}}, \label{ion-pado}
\end{equation} 
where $\phi$ accounts for secondary ionization (i.e., ionization events induced 
by secondary electrons released during primary ionization events), 
$\mathrm{d}^3N_{\mathrm p}/\big(\mathrm{d}E_{\mathrm{p}}\,{\mathrm d}A\,{\mathrm d}t\big)$ is the differential 
CR proton flux, and $\sigma_{{\mathrm{ion}}}^{{\mathrm H}_2}$ is the direct ionization cross-section of molecular 
hydrogen by protons. 
The primary CR proton flux 
can be obtained with the help of gamma-ray observations and the differential CR proton number density (\ref{eq:sol_transp_eq}). 
Treating the observed gamma-ray fluxes from SNR-MC systems as isotropic emission from the regions enclosed by the SNR shock fronts, 
the primary CR proton fluxes in these regions are constructed, 
assuming that the gamma-ray emission is a result of the formation and subsequent decay of neutral pions from interactions of 
the CR protons with ambient hydrogen, such that the corresponding gamma-ray fluxes match the observational data. This is done by 
means of the proton-proton 
interaction model given in \cite{kelner2006}, which 
describes gamma-ray emission for primary energies above 100~GeV, and with the delta distribution 
approximation for the pion-production cross-section at lower energies \citep[see, e.g.,][]{mannheim1994, aharonian2000}. 
In order to study the transport of these CR protons into 
the interior regions of the clouds, one uses the CR proton fluxes outside the MCs as initial boundary values for the 
differential CR proton number densities (\ref{eq:sol_transp_eq}) and determines the 
function $Q_p(p)$ and the normalization constant $Q_{\mathrm{norm}}$ in Eq.\ 
(\ref{source}) 
for the objects under consideration. 
To this end, the CR proton fluxes outside the MCs derived from gamma-ray observations are expressed in terms of a (dimensionless) 
spectral shape function $\Phi_{\mathrm{p}}(E_{\mathrm{p}})$ and a normalization constant $a_{\mathrm{p}}$, and 
linked to the momentum source factor of Eq.\ (\ref{source}) and $Q_{\mathrm{norm}}$ by simple multiplication with the CR proton 
kinetic energy $E_{\mathrm{p}}$ and division by the particle speed used in the Kelner gamma-ray emission model, the 
speed of light $c$, yielding 
\begin{equation}
 Q_{\mathrm{norm}}\cdot Q_p(p) = a_{\mathrm{p}} \cdot \Phi_{\mathrm{p}}(E_{\mathrm{p}}) \cdot \frac{E_{\mathrm{p}}(p)}{c} = 
 \bigg(\frac{\mathrm{d}^3N_{\mathrm p}}{\mathrm{d}E_{\mathrm{p}}\,{\mathrm d}A\,{\mathrm d}t}\bigg)_{\mathrm{obs}} \cdot \frac{E_{\mathrm{p}}(p)}{c}. 
\end{equation}
The spectral shape function $\Phi_{\mathrm{p}}(E_{\mathrm{p}})$ that is used to model the observed 
gamma-ray emission as described in more detail in \cite{schuppan2012} is usually given by 
a set of typical expressions for SNRs detected at gamma-ray energies: 
\begin{enumerate}[(a)]
 \item a single power-law in the kinetic energy $E_{\mathrm{p}} = \big(p^2c^2 + m^2_{\mathrm p} c^4\big)^{1/2}-m_{\mathrm p} c^2$ \\
       $\Phi_{\mathrm{p}}(E_{\mathrm{p}})= (E_{\mathrm{p}}/E_{\mathrm{norm}})^{-\alpha},$
 \item a single power-law in the kinetic energy with an exponential cut-off \\
       $\Phi_{\mathrm{p}}(E_{\mathrm{p}})= (E_{\mathrm{p}}/E_{\mathrm{norm}})^{-\alpha}\cdot\exp\left(-E_{\mathrm{p}}/E_\mathrm{{c}}\right),$
 \item a double power-law in the kinetic energy \\
       $\Phi_{\mathrm{p}}(E_{\mathrm{p}})= (E_{\mathrm{p}}/E_{\mathrm{norm}})^{-\alpha}\cdot\left(1+E_{\mathrm{p}}/E_{\mathrm{br}}\right)^{\alpha-\alpha_{\mathrm h}},$
 \item a double power-law in the kinetic energy with an exponential cut-off \\
       $\Phi_{\mathrm{p}}(E_{\mathrm{p}})= (E_{\mathrm{p}}/E_{\mathrm{norm}})^{-\alpha}\cdot\left(1+E_{\mathrm{p}}/E_{\mathrm{br}}\right)^{\alpha-\alpha_{\mathrm h}}
        \cdot\exp\left(-E_{\mathrm{p}}/E_\mathrm{{c}}\right),$
\end{enumerate}
where $E_{\mathrm{norm}}$, $E_\mathrm{c}$, and $E_{\mathrm{br}}$ are normalization, cut-off, and break energies, respectively, 
$\alpha$ is the effective power-law index for energies below the break energy, and $\alpha_{\mathrm h}$ is the effective power-law 
index above the break energy. 
Matching the different expressions for the CR proton fluxes to the observed gamma-ray emission of the four specific objects 
considered here via the Kelner model reveals that for W49B and W44, 
the description (c) and for 3C~391 and CTB~37A the description (b) reproduces the 
observational data best. The corresponding best-fit parameters are shown in Table \ref{tab_p_param}. 
Note that in order to obtain values for the CR proton flux normalization constant $a_{\mathrm{p}}$, one 
has to make assumptions regarding the volumes $V = 4\pi R^3 f_{\mathrm{V}}/3$ (to calculate the edge length 
$l_{\mathrm{c}}$ of the cubic emission regions), where $f_{\mathrm{V}}$ is a volume-filling 
factor (i.e., the homogeneous fraction (with hydrogen density n$_{\rm H}$) of the volume of the SNR-MC complex 
where the pions are formed, cf.\ Table \ref{tab_p_param}), 
and $R$ is the radius of the SNR shock front (cf.\ Table \ref{tab_dist_ext}), as well as 
the hydrogen densities of these emission regions. 
Here, emission volumes of $V=21.4\,$pc$^{3}$ for W49B, $V=1.88\cdot10^{4}\,$pc$^{3}$ for W44, $V=8.38\cdot10^{3}\,$pc$^{3}$ 
for 3C~391, and $V=3.99\cdot10^{4}\,$pc$^{3}$ for CTB~37A, and homogeneous hydrogen densities of $n_{\mathrm H}=100$ cm$^{-3}$ are used. 
\begin{table}[b]
\centering{
\begin{tabular}{cccccccc}
  \hline\hline \\ [-10pt]
  object & $a_{\mathrm{p}}$ [erg$^{-1}$ s$^{-1}$ cm$^{-2}$] & $\alpha $ & $E_{\mathrm{br}}$ [GeV] & $\alpha_{\mathrm h}$ & $E_\mathrm{{c}}$ [GeV] & age [kyr] & $f_{\mathrm{V}}$\\
  \hline
  W49B & 1.10$\cdot$10$^9$ & 2.0 & 4 & 2.7 & - & 1 & 0.06 \\
  %\hline
  W44 & 8.3$\cdot$10$^5$ & 1.74 & 9 & 3.7 & - & 20 & 1 \\
  %\hline
  3C 391 & 4.6$\cdot$10$^6$ & 2.4 & - & - & 100,000 & 4 & 1 \\
  %\hline
  CTB 37A & 3.2$\cdot$10$^4$ & 1.7 & - &  - & 80,000 & 2 & 1 \\
  \hline 
\end{tabular} 
\vspace{0.5cm}
\caption{Spectral parameters of the CR proton sources used to match the observed gamma-ray emission. For all 
objects, $E_{\mathrm{norm}}=1\,$GeV, $D(p)=10^{28}\cdot\left(p/(m_{\mathrm p}c)\right)^{4/3}$ cm$^2$ s$^{-1}$, and the 
hydrogen density of the MC was assumed to be homogeneously distributed with a value of $n_{\mathrm H}=100$ cm$^{-3}$. 
For W49B, a volume-filling factor of $0.06$ was used \citep{abdo(W49B)2010}. Since for the other objects no 
solid observational estimates of the volume-filling factors exist, volume-filling factors of unity were used.\label{tab_p_param}}}
\end{table}
Since the break energies and the cut-off energies of all these objects are large compared to 
the maximum kinetic energy considered, only 
the single power-law (a), which is the $E_{\mathrm{p}}/E_{\mathrm{br}}\rightarrow0$ limit and the 
$E_{\mathrm{p}}/E_{\mathrm{c}}\rightarrow0$ limit of the parametric descriptions (b)-(d), is applied in calculating the ionization rates. 
After having determined the source term (\ref{source}) for each object, the solution of the 
transport equation, the CR proton number density (\ref{eq:sol_transp_eq}), is used in order to account for 
the propagation of the CR protons into the MCs and, thus, to calculate the ionization rates as functions 
of the penetration depth into the cloud as follows: 
the CR proton fluxes inside the MCs are related to the corresponding differential CR proton number densities by derivating the CR 
proton number density with respect to the CR proton kinetic energy and by multiplication with the effective particle speed $v$ 
\begin{equation}
 \frac{{\mathrm d}^3N_{\mathrm p}\big(\vec{r},p(E_{\mathrm{p}}),t\big)}{\mathrm{d}E_{\mathrm{p}}\,{\mathrm d}A\,{\mathrm d}t} = 
 \frac{\mathrm{d} n_{\mathrm p}\big(\vec{r},p(E_{\mathrm{p}}),t\big)}{\mathrm{d} E_{\mathrm{p}}}\cdot v.
\end{equation}
The effective particle speed is a combination of the diffusion speed $v_{\mathrm{diff}} = \Delta s/T_{\mathrm{diff}} = 2 D(p)/\Delta s$, 
where $\Delta s$ denotes the distance over which a particle was diffusing and $T_{\mathrm{diff}}=(\Delta s)^2/(2\,D(p))$ is the corresponding 
diffusion timescale, and the relativistic particle speed, $v_{\mathrm{rel}}=p\,c/\big(p^2+m_{\mathrm{p}}^2c^2\big)^{1/2}$, which is 
determined by initial values at the collision regions of the shock fronts and the MCs. 
If one of these speeds is dominant, it 
can be chosen as an approximation of the effective particle speed. It is shown in Sect.\ \ref{results} that 
$v_{\mathrm{diff}}$ dominates over $v_{\mathrm{rel}}$ at all relevant penetration depths for the specific assumed diffusion coefficient. 
Here, in order to circumvent the problem that the time that has passed since the injection of the CR particles 
into the clouds until the measurement, while restricted to $0<t<t_{\mathrm{age}}$, is generally not known, 
one couples the propagation time $t$ to the position and the momentum of the particles via 
$t = T_{\mathrm{diff}}(|\vec{r}|,p) = |\vec{r}|^2/(2\,D(p))$, 
which is the mean time for a particle with momentum $p$ to propagate the distance $|\vec{r}|$ in the case of 
standard three-dimensional Gaussian diffusion, or via $t = T_{\mathrm{rel}}(|\vec{r}|,p) = |\vec{r}|/v_{\mathrm{rel}}(p)$. 
Moreover, the ionization rate is evaluated only along the $z$-axis (cf.\ Sect.\ \ref{TE}). 
The primary CR proton flux, as a function of this specific penetration depth into the MC at time $t=T(z,p)$ after the injection 
reads 
\begin{equation}
 \frac{{\mathrm d}^3 N_{\mathrm p}\big(z,p(E_{\mathrm{p}})\big)}{\mathrm{d}E_{\mathrm{p}}\,{\mathrm d}A\,\mathrm{d}t} 
 = \frac{\mathrm{d} n_{\mathrm p}\big(x=y=0, z, p(E_{\mathrm{p}}), t=T(z,p(E_{\mathrm{p}}))\big)}{\mathrm{d} E_{\mathrm{p}}} 
 \cdot v \label{spec_x_y_t}
\end{equation}
with $T=T_{\mathrm{diff}}$ or $T=T_{\mathrm{rel}}$ and $v=v_{\mathrm{diff}}$ or $v=v_{\mathrm{rel}}$, respectively. 
Note that the ionization rate at a certain penetration depth is not generated instantaneously, but builds up when the CR particles 
of the highest momentum considered have had the time to reach this penetration depth with their effective speed, and saturates once 
the CR particles of the lowest momentum arrive. 
This implies that the ionization rate should only be calculated up to the penetration depth 
$z_{\max,\mathrm{diff}} = \sqrt{2\,D(p_{\min})\, t_{\mathrm{age}}}$\, or $z_{\max,\mathrm{rel}} = v_{\mathrm{rel}}(p_{\min})/t_{\mathrm{age}}$, 
respectively, which correspond to the distances the particles with minimum momentum move within the ages of the 
SNRs. 

After having computed the constrained CR proton fluxes (\ref{spec_x_y_t}) for the MCs, 
the ionization rate of molecular hydrogen as a function of the penetration 
depth into the cloud $z$ can, therefore, be represented as the following integral 
\begin{align}
 \zeta^{{\mathrm H}_2}(z) = \left(1 +\phi\right)\int_{p_{\min}}^{p_{\max}} 
 \frac{\mathrm{d} n_{\mathrm p}\big(z, p(E_{\mathrm{p}}), t=T(z,p(E_{\mathrm{p}}))\big)}{\mathrm{d} E_{\mathrm{p}}} \cdot v\big(p) \cdot 
 \sigma_{{\mathrm{ion}}}^{{\mathrm H}_2}(p) \,{\mathrm d}p, \,\,\,\,\,\, \mathrm{where} \,\, 0\le z\le z_{\max}. \label{ion_prof_full}
\end{align}
The cross-section for direct ionization of molecular hydrogen by protons, $\sigma_{{\mathrm{ion}}}^{{\mathrm H}_2}(p)$, 
can be given via the parametric expression \citep{rudd1985} 
\begin{equation}
 \sigma_{{\mathrm{ion}}}^{{\mathrm H}_2}(p) = \frac{\sigma_{\mathrm{l}}(p)\sigma_{\mathrm{h}}(p)}{\sigma_{\mathrm{l}}(p)+\sigma_{\mathrm{h}}(p)}
\end{equation}
with the low-energy component $\sigma_{\mathrm{l}}=4\pi\,a_0^2\,C\,x(p)^D$ and the high-energy component 
$\sigma_{\mathrm{h}}=4\pi\,a_0^2[A \cdot \ln(x(p)+1) + B]$ \\
\noindent $\cdot x(p)^{-1}$, where 
$x(p)=\left(m_{\mathrm e}c^2/I_{\mathrm H}\right)\left(\big[1 + \big(p/(m_{\mathrm p}c)\big)^2\big]^{1/2}-1\right)$, 
$a_0=5.29\cdot10^{-9}$\,cm is the Bohr radius, the parameters $A=0.71$, $B=1.63$, $C=0.51$, as well as $D=1.24$, $m_{\mathrm e}$ 
and $m_{\mathrm p}$ are the electron and proton masses, respectively, and $I_{\mathrm{H}}=13.598\,\mathrm{eV}$ is the 
ionization potential of atomic hydrogen. 
The integration boundaries $p_{\min} = 0.15\,m_{\mathrm{p}}c$ and $p_{\max} = 0.86\,m_{\mathrm{p}}c$ correspond to 
the minimum and maximum CR proton energies $E_{\mathrm{p},\min}=10\,\mathrm{MeV}$ and \mbox{$E_{\mathrm{p},\max}=280\,\mathrm{MeV}$.} 
In general, the dimensionless quantity $\phi$, which accounts for secondary ionization, depends 
on the energies of the primary CR protons, but for the energy range of the protons covered in this work, 
$\phi=0.6$ is an adequate constant approximation \citep{cravens1978}. 
The secondary electrons have, on average, rather low kinetic energies that they lose rapidly 
in interactions with matter, electromagnetic radiation and the magnetic fields inside the MCs. This 
allows to consider the effect of secondary ionization as an instantaneous phenomenon, i.e., to neglect the 
propagation of the secondary electrons.
The large number of primary CR particles moreover justifies the treatment of the corresponding secondary 
ionization via a constant. 
The integral (\ref{ion_prof_full}) is solved numerically for each object, using the adaptive Monte Carlo
algorithm VEGAS introduced by \cite{lepage1978}. While in standard Monte Carlo routines the sample points are chosen 
according to a preassigned distribution function, VEGAS iteratively concentrates, in an adaptive manner, 
the distribution of the sample points on those intervals where the integrand contributes the most, starting 
from a uniform distribution of the sample points. This algorithm is slower than, e.g., a Gauss-Kronrod quadrature, 
but is more reliable and stable, particularly for rapidly changing integrands in one-dimensional integrations, 
which are performed here. 
The VEGAS routine is implemented as a C{}\verb!++! algorithm via the GNU Scientific Library \citep{galassi2009}. 
The results of these integrations for $v=v_{\mathrm{diff}}$ and $t=T_{\mathrm{diff}}$ as well as $v=v_{\mathrm{rel}}$ and 
$t=T_{\mathrm{rel}}$ are presented in Sect.\ \ref{results}. 

%================================================
\section{Photoionization profiles\label{X-ray}}
%================================================
In order to determine whether there is a dominant process generating the ionization signatures of SNR-MC systems, one has to 
derive the ionization profiles for all eligible processes. To this end, the profiles emerging from photoionization are studied 
in addition to those induced by CR protons. 

Photoionization is primarily induced by Extreme UV (EUV) radiation ($13.6\,\rm{eV}\le E\le200\,$eV) and 
X-rays ($E>200\,$eV). EUV radiation is 
quite effective in ionizing molecular hydrogen, but its penetration depth into MCs is short 
because it is rapidly absorbed. 
X-rays, however, are capable of traversing larger columns of matter, with their soft (low-energy) component 
losing energy faster than their hard (high-energy) component. In fact, hard X-rays might penetrate even 
farther into MCs than CRs \citep{tielens2010}. 
In this section, the X-ray fluxes in the SNR-MC systems W49B, W44, 3C~391, and CTB~37A are derived 
at the MC surfaces using observational data, assuming that the fluxes are composed of synchrotron radiation of CR electrons 
emitted at the SNR shock fronts, and for the transport into the 
MCs via the Beer-Lambert law. From these fluxes, the photoinduced ionization 
rates are calculated as functions of the penetration depth, and compared with those induced by CRs.

%===============================================
\subsection{Photon fluxes at the MC surfaces\label{X_spec}}
\begin{figure}[ht]%
\centering
\subfigure[][]{%
\label{W49B_fit_X}%
\includegraphics[width=0.48\columnwidth]{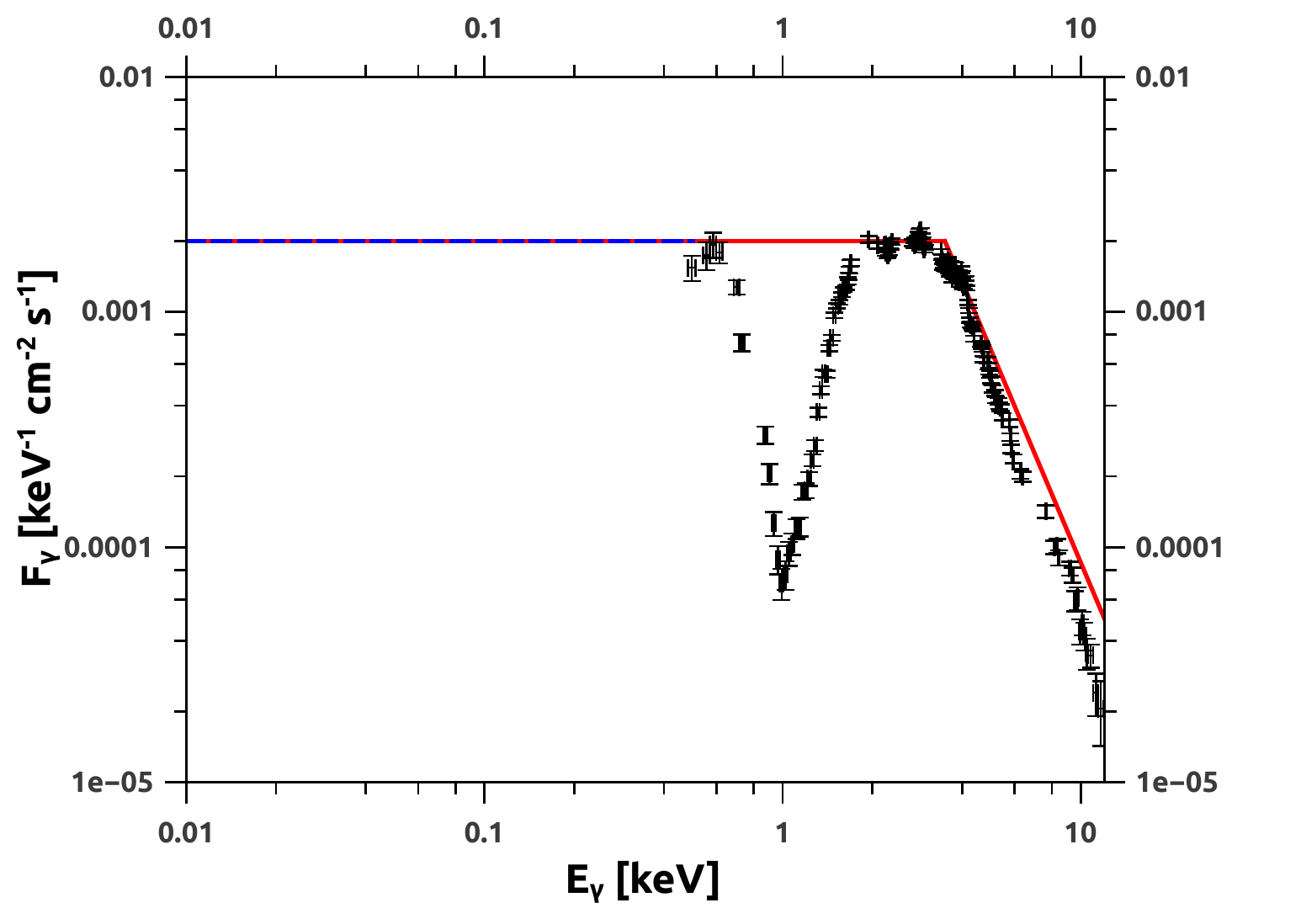}}%
\hspace{8pt}%
\subfigure[][]{%
\label{W44_fit_X}%
\includegraphics[width=0.48\columnwidth]{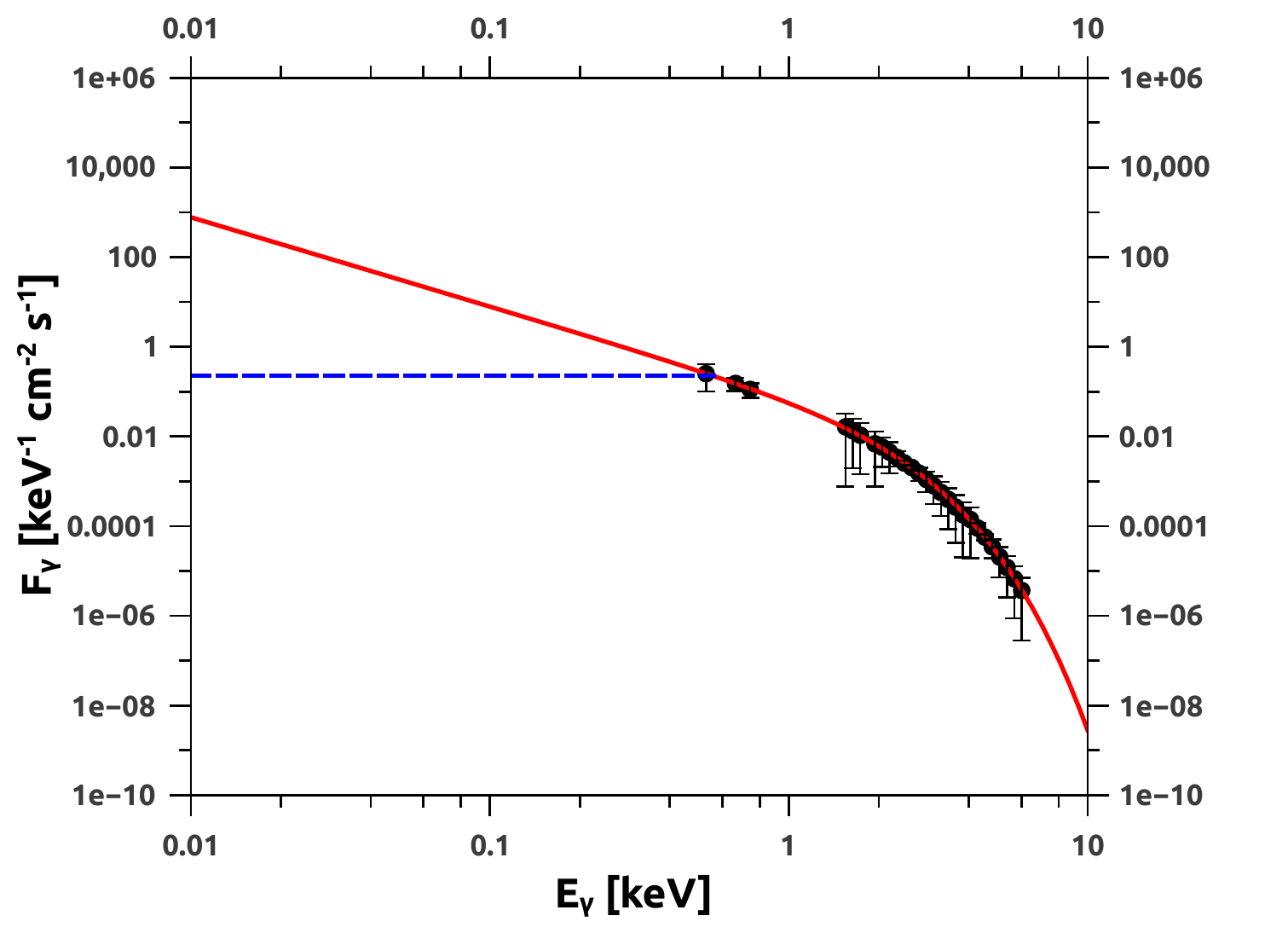}}\\
\subfigure[][]{%
\label{3C391_fit_X}%
\includegraphics[width=0.48\columnwidth]{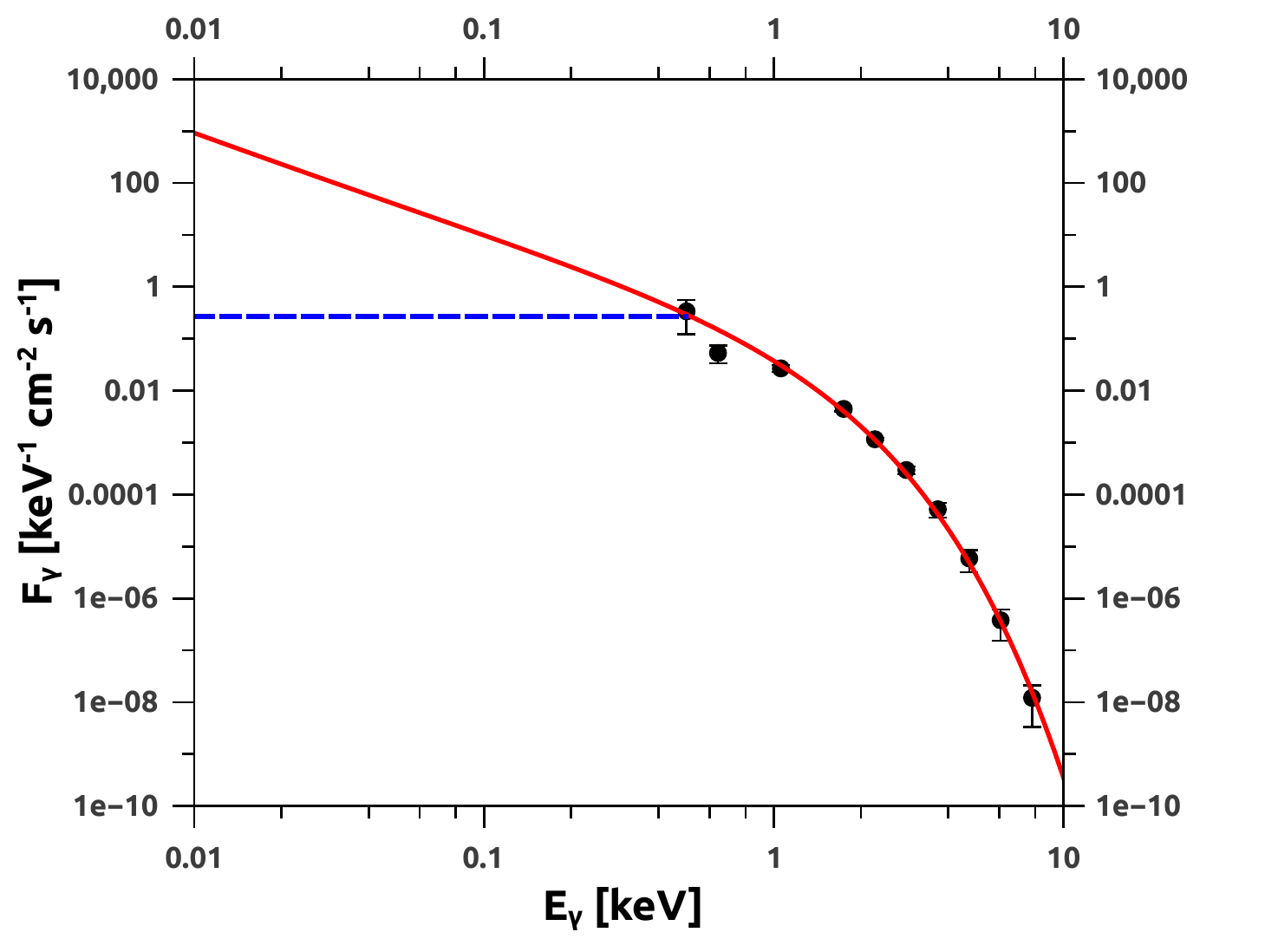}}%
\hspace{8pt}%
\subfigure[][]{%
\label{CTB37A_fit_X}%
\includegraphics[width=0.48\columnwidth]{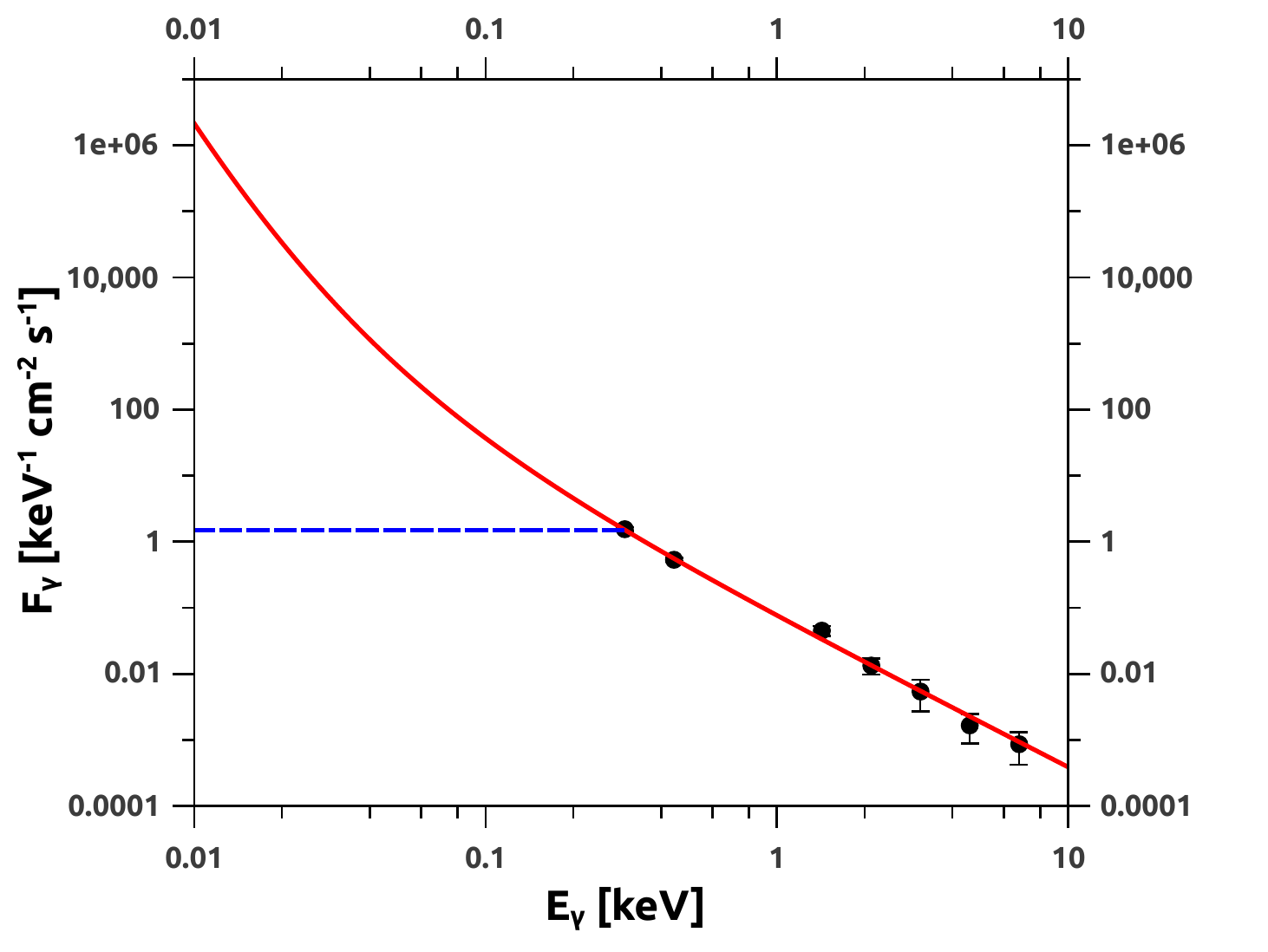}}%
\caption[...]
 {Parametric models (I) (red, solid lines), (II) (blue, long-dashed lines), and (III) fitted to X-ray flux data 
 samples (black dots and error bars) for the objects W49B \subref{W49B_fit_X}, W44 \subref{W44_fit_X}, 3C~391 
 \subref{3C391_fit_X}, and CTB~37A \subref{CTB37A_fit_X}. The X-ray data samples used are taken from \cite{ozawa2009} 
 (Suzaku) for \subref{W49B_fit_X}, \cite{harrus2006} (XMM-Newton) for \subref{W44_fit_X}, \cite{chen2001} (ASCA) 
 for \subref{3C391_fit_X}, and \cite{sezer2011} (Suzaku) for \subref{CTB37A_fit_X}.}%
\label{fits_X}%
\end{figure}

%===============================================
The X-ray fluxes in the vicinity of MCs are derived assuming isotropic radiation emission from the 
SNR shock fronts, which are known to emit synchrotron radiation of CR electrons in the form of X-rays. It is furthermore assumed that the SNR 
shock fronts coincide with the surfaces of the clouds, in accordance with the geometry used for the calculation of the ionization 
induced by CR protons. 
Note that X-ray observations of SNRs associated with MCs provide near-Earth X-ray flux data, which is 
inadequate for the derivation of the X-ray fluxes at the MCs because, on the one hand, 
there are energy-dependent losses from interactions of the photons with the interstellar medium in between the emission region 
and the detector, and on the other hand, there is a geometrical 1/$d^2$-attenuation of the photon fluxes 
from the source along the path toward the instrument. 
Thus, the X-ray fluxes at the cloud surfaces, $F_{\mathrm{X\textnormal{-}ray,\,cs}}$, can be derived from the X-ray 
fluxes at the instrument, $F_{\mathrm{X\textnormal{-}ray,\,in}}$, by 
\begin{equation}
 F_{\mathrm{X\textnormal{-}ray,\,cs}} = F_{\mathrm{X\textnormal{-}ray,\,in}} \cdot \left(\frac{d}{R}\right)^{2},
 \label{X_surf}
\end{equation}
where $d$ is the distance between the SNR-MC system and Earth, and $R$ is the radial extension of the SNR shock front. The values 
of the parameters $d$ and $R$ for the objects considered are given in Table~\ref{tab_dist_ext}. 
\begin{table}[htb]
\centering{
\begin{tabular}{ccc}
	\hline\hline \\ [-10pt]
  object & $d$ [kpc] & $R$ [pc] \\ 
  \hline \\ [-10pt]
  W49B & 8.0$^{+1.2}_{-0.4}$ & 4.4$^{+0.7}_{-0.2}$ \\
  %\hline
  W44 & 3.3${\pm0.4}$ & 16.5$\pm2.0$ \\
  %\hline
  3C 391 & 7.2${\pm0.3}$ & 12.6$\pm0.5$ \\
  %\hline
  CTB 37A & 8.0$^{+1.5}_{-1.7}$ & 21.2$^{+4.0}_{-4.5}$ \\
  \hline
\end{tabular}
\vspace{0.5cm}
\caption{SNR-MC system-Earth distances and radial SNR shock extensions.}\label{tab_dist_ext}}
\end{table}
The attenuation of the X-ray fluxes from interactions with traversed matter between the emission region and the 
detector is already taken into account in the data samples of the X-ray observations $F_{\mathrm{X\textnormal{-}ray,\,in}}$. 
Since the photoinduced ionization rate is given, as in the case of CR-induced ionization (cf.\ Eq.\ (\ref{ion-pado})), 
in terms of an integral expression that includes the photon flux as a function of the photon energy, a continuous description 
of the observational data 
is required. This is achieved by fitting a continuous parametric model function for the X-ray fluxes to the observational 
data samples. The parametric model function for the photon fluxes of two of the four SNR-MC systems studied here, 
W44 and 3C~391, is chosen to be a triple power-law 
\begin{equation}
 F_{\mathrm{X\textnormal{-}ray,\,in}}(E) = F_0\cdot\left(\frac{E}{1\,\mathrm{keV}}\right)^{-a}\cdot\left(1+\frac{E}{b_1}\right)^{-a_1}\cdot\left(1+\frac{E}{b_2}\right)^{-a_2},
\end{equation}
where $a$, $a_1$, and $a_2$ are dimensionless spectral indices, $b_1$ and $b_2$ denote break energies, respectively, 
and $F_0$ is a constant. 
Owing to their significantly different X-ray data samples, different parametric models for CTB~37A and W49B are used. 
These are, for CTB~37A, 
\begin{equation}
 F_{\mathrm{X\textnormal{-}ray,\,in}}(E) = F_0\cdot\left(\frac{E}{1\,\mathrm{keV}}\right)^{-a}\cdot\left(1+\frac{E}{b_1}\right)^{-a_1},
\end{equation}
and for W49B, 
\begin{equation}
 F_{\mathrm{X\textnormal{-}ray,\,in}}(E) = \begin{cases}
                                             F_0 &\,\,\,\mathrm{for}\,\,\, E<3.5\,\mathrm{keV}\\
                                             F_0 \cdot {\displaystyle \left(\frac{E}{3.5\,\mathrm{keV}}\right)^{-a_1}} &\,\,\,\mathrm{for}\,\,\, E\ge3.5\,\mathrm{keV}.
                                           \end{cases}
\end{equation}
Applying these simpler parametric descriptions leads to an improved $\chi^2$-value per degree of freedom since a lesser number 
of fitting parameters is used. 
The best-fit parameters are given in Table~\ref{tab_X_surf}. 
The extremely low value of $\chi^2$ per degree of freedom for the object W44 results from the large error bars of 
the X-ray data, cf.\ Fig.\ \ref{W44_fit_X}. 
\begin{table}[ht]
\centering{
\begin{tabular}{cccccccc}
	\hline\hline \\ [-10pt]
  object & $F_0$\,[s$^{-1}$\,cm$^{-2}$\,keV$^{-1}$] & $a$ & $b_1$\,[keV] & $a_1$ & $b_2$\,[keV] & $a_2$ & $\chi^2$/d.o.f.\\
  \hline
  W49B & 0.00200 & 0.0 & 3.5 & 3.0 & - & - & - \\
  %\hline
  W44 & 0.082 & 1.99 & 2.70 & -6.0 & 97.0 & 223.0 & 0.0068 \\
  %\hline
  3C 391 & 0.066 & 2.06 & 0.050 & -0.40 & 50.0 & 90.0 & 1.11 \\
  %\hline
  CTB 37A & $1.64\cdot10^{-13}$ & 8.9 & 0.0181 & -6.7 & - & - & 1.12 \\
  \hline
\end{tabular}
\vspace{0.5cm}
\caption{Best-fit parameters of the X-ray flux models and their $\chi^2$ per degree of freedom.}\label{tab_X_surf}}
\end{table}
For the objects W44, 3C~391, and CTB~37A, certain X-ray data points were excluded because they are 
caused by line emission originating along the path between the near-Earth measuring instrument and the SNR-MC systems 
\citep{ozawa2009, harrus2006, chen2001, sezer2011}. 

The fitting of the parametric models to the observational X-ray data samples was performed by minimizing $\chi^2$ per degree 
of freedom based on the Nelder-Mead method \citep{NelderMead1965}. 
This method is a non-linear optimization technique that does not require the derivatives of $\chi^2$ in order to 
minimize $\chi^2$ with respect to the optimization parameters. It is particularly useful in multi-dimensional parameter space 
optimization. 
This routine is implemented as a C{}\verb!++! code via the GNU Scientific Library. 

Since the total photoabsorption cross-section, which is relevant for the calculation of photoionization, is highest 
at low photon energies, the treatment of the EUV domain is crucial 
and consequently has to be included in this study. To evaluate the influence of low-energy photons on the photoionization rate, 
three descriptions of the low-energy photon fluxes are considered: 
\begin{enumerate}[(I)]
 \item extrapolation of the parametric model function used to fit the observed X-ray flux values toward lower photon energies, 
 \item a constant photon flux equal to the value of the parametric model function at the lowest photon energy covered by the observational data, 
 \item a hard cut-off below the lowest photon energy covered by the observational data, that is, no extrapolation, 
\end{enumerate}
where the lowest photon energies for which X-ray data are available are $0.5\,$keV for W49B as well as 3C~391, $0.56\,$keV for 
W44, and $0.3\,$keV for CTB~37A. 
The parametric models and the different descriptions of the low-energy photon fluxes are shown along with the observational X-ray 
data samples in Fig.\ \ref{fits_X}. 

%===============================================
\subsection{Photon fluxes inside MCs and photoinduced ionization profiles \label{X_att}}
%===============================================
Using the photon fluxes obtained at the surfaces of the MCs, the corresponding 
fluxes inside the clouds, which undergo an attenuation due to interactions with % traversed 
cloud matter, as functions of the photon 
energy and the penetration depth, are calculated. This is done by means of the Beer-Lambert formula, 
\begin{equation}
 F_{\mathrm{X\textnormal{-}ray}}(z,E) =  F_{\mathrm{X\textnormal{-}ray,\,cs}}\cdot\exp\big(-\tau(z,E)\big)\,,
\end{equation}
which relates the photon flux $F_{\mathrm{X\textnormal{-}ray}}$ at a certain penetration depth and energy to the initial photon flux 
(\ref{X_surf}) at the cloud surface. The quantity 
$\tau(z,E) = \sigma_{\mathrm{pa}}(E)\, n_{\mathrm{H}}\, z$ is the optical depth for a homogeneous 
cloud, $\sigma_{\mathrm{pa}}(E)$ is the total photoabsorption cross-section, and $n_{\mathrm{H}}$ is the 
hydrogen density. 
It should be remarked that the actual photon paths are longer than the direct ones, for which deviations caused by scattering 
are neglected. Therefore, the photoinduced ionization 
signatures calculated here are upper limits, leading to conservative estimates with respect to the 
detectability of CR-induced ionization signatures, since under the influence of scattering, the photons lose 
more energy at lower penetration depths, which leaves less energy for ionization at large penetration depths. 
The photoinduced ionization rate, as a function of the penetration depth, following \cite{maloney1996}, yields 
\begin{equation}
 \zeta^{{\mathrm H}_2}_{\mathrm{X\textnormal{-}ray}}(z) = \frac{f_{\mathrm{i}}}{I_{\mathrm{H_2}}}\int_{E_{\min}}^{E_{\max}} 
 F_{\mathrm{X\textnormal{-}ray}}(z,E)\cdot E \cdot\sigma_{\mathrm{pa}}(E)\,\mathrm{d}E, \label{photoion}
\end{equation}
where $f_{\mathrm{i}}$ is the fraction of the absorbed photon energy leading to ionization, 
$I_{\mathrm{H_2}}=15.4\,$eV is the ionization potential of molecular hydrogen, and the integral 
is the total energy deposition by the absorbed photons per hydrogen nucleus. 
The parametrization used for a description of the total photoabsorption cross-section per hydrogen nucleus is also provided 
in \cite{maloney1996} as an empirical, broken power-law fit to experimental data 
\begin{align}
 \sigma_{\mathrm{pa}}(E) &= \sigma_0 \cdot\left(\frac{E}{1\,\mathrm{keV}}\right)^{-\gamma}\\ 
 \nonumber \\ %\vspace{1.0cm}
 \mathrm{with} \,\,\, \sigma_0 &= 6.0 \cdot 10^{-23}\,\mathrm{cm^2}\,\,\,\mathrm{and}\,\,\, \gamma = 3\,\,\,\mathrm{for}\,\,\, 0.014\,\mathrm{keV}<E\le0.5\,\mathrm{keV},& \nonumber \\
 \sigma_0 &= 2.6 \cdot 10^{-22}\,\mathrm{cm^2}\,\,\,\mathrm{and}\,\,\, \gamma = 8/3\,\,\,\mathrm{for}\,\,\, 0.5\,\mathrm{keV}<E\le7.0\,\mathrm{keV},& \nonumber \\
 \sigma_0 &= 4.4 \cdot 10^{-22}\,\mathrm{cm^2}\,\,\,\mathrm{and}\,\,\, \gamma = 8/3\,\,\,\mathrm{for}\,\,\, 7.0\,\mathrm{keV}<E\,. \nonumber 
\end{align}
The breaks in the above expression come from contributions of heavy elements in the MCs to the total 
cross-section for which there is a certain threshold energy for shell absorption, namely oxygen $K$-shell 
absorption at 0.5~keV and iron $K$-shell absorption at 7~keV. 
In contrast to the ionization rate induced by cosmic rays, the photoinduced ionization rate is not dominated by direct ionization events. 
The fraction of the absorbed photon energy yielding ionization is approximately 
$f_{\mathrm{i}}\approx40\%$ of the deposed photon energy \citep{voit1991,dalgarno1999}, so a photon with $E=1\,$keV 
absorbed by the cloud medium induces a total of 26 ionization events. This allows a description of the total 
photoionization rate in terms of the photon energy deposed in the cloud matter. The lower integration boundary is set 
at the photon energy required to produce an ion pair, $E_{\min}=37\,$eV, in order to obtain an upper limit 
of photoinduced ionization, but it should be noted that no X-ray measurements for $E\lesssim0.1\,$keV exist. The 
upper integration boundary is fixed at 10\,keV, but a higher value would not change the results 
significantly because of the rapid decrease of both the total photoelectric energy deposition cross-section and the photon flux 
with increasing photon energy. 
The ionization profiles induced by photons (\ref{photoion}) are calculated numerically for each object as well as each parametric description 
of the low-energy photon flux (I)-(III) using the VEGAS integration technique and shown in the following section.

%================================================
\section{Results\label{results}}
\begin{figure}[ht]%
\centering
\subfigure[][]{%
\label{W49B_dyn_diff}%
\includegraphics[width=0.48\columnwidth]{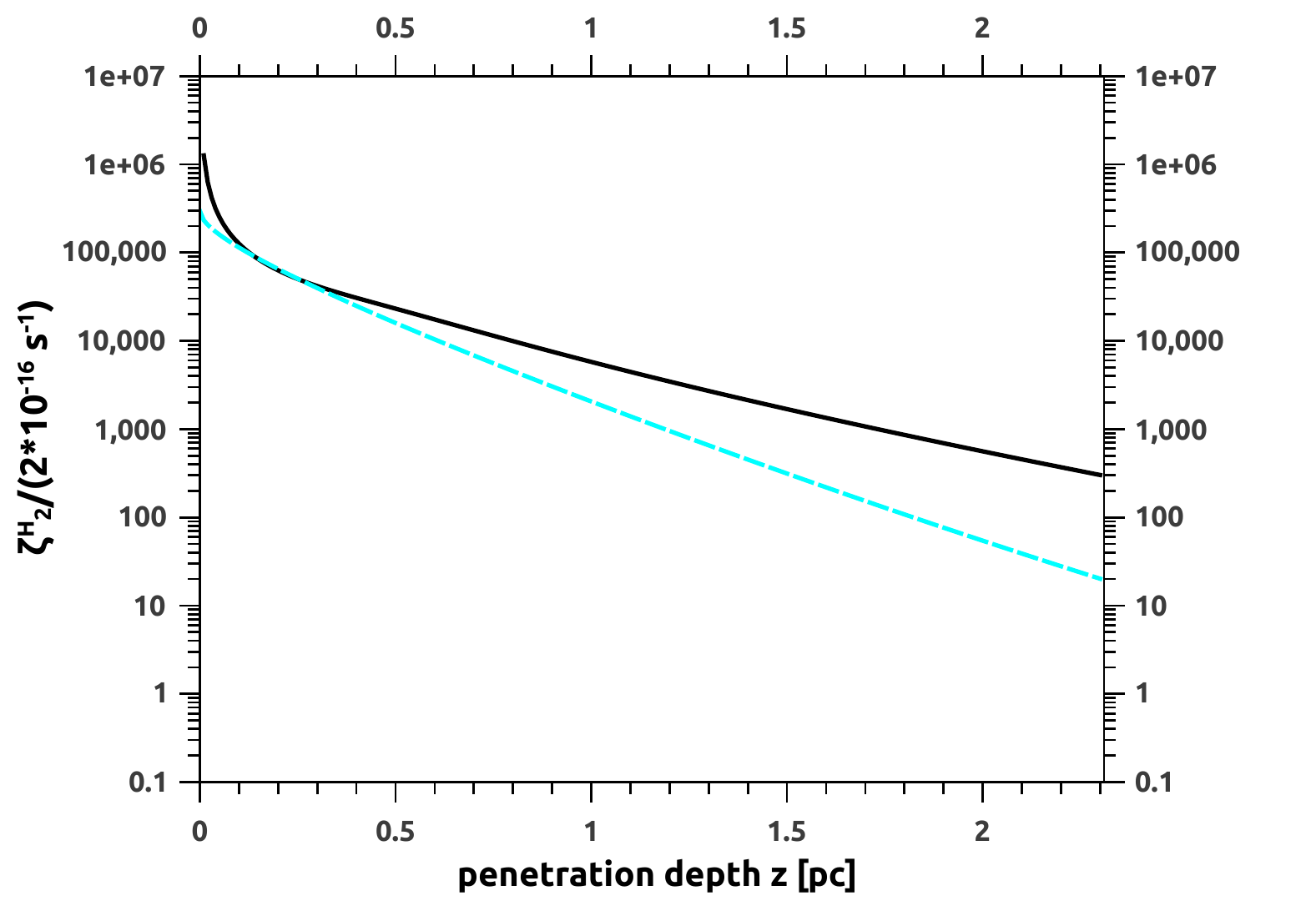}}%
\hspace{8pt}%
\subfigure[][]{%
\label{W44_dyn_diff}%
\includegraphics[width=0.48\columnwidth]{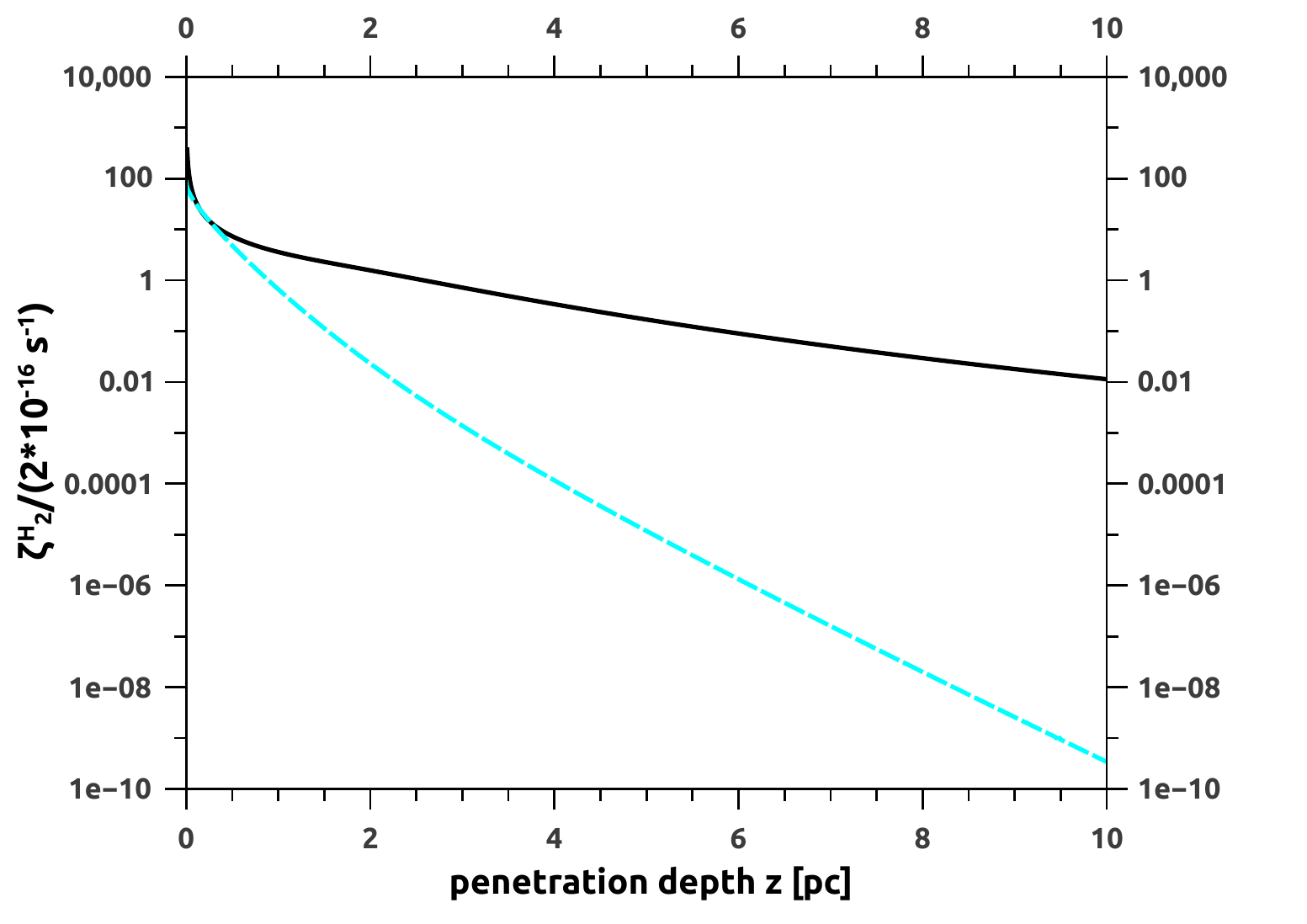}}\\
\subfigure[][]{%
\label{3C391_dyn_diff}%
\includegraphics[width=0.48\columnwidth]{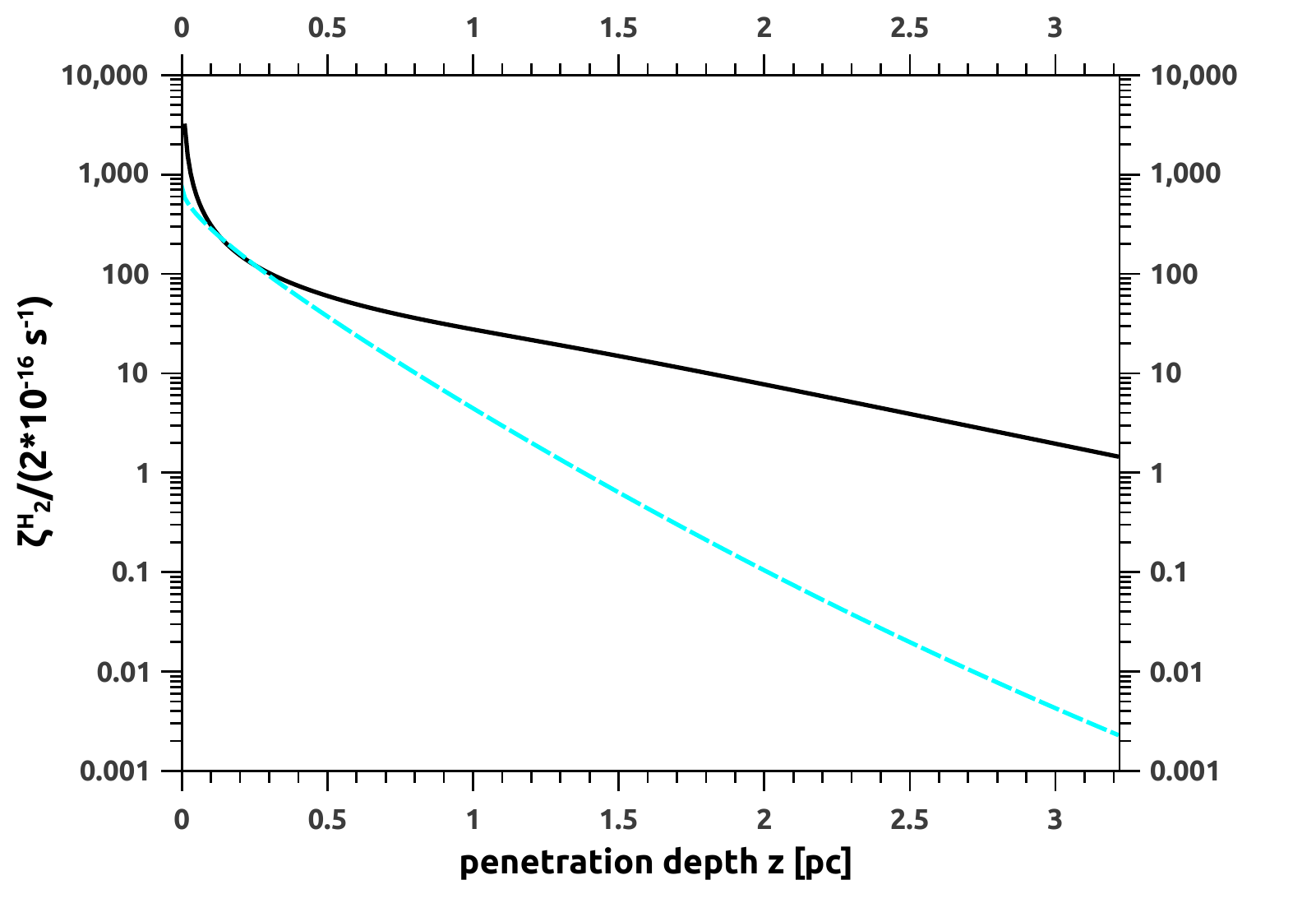}}%
\hspace{8pt}%
\subfigure[][]{%
\label{CTB37A_dyn_diff}%
\includegraphics[width=0.48\columnwidth]{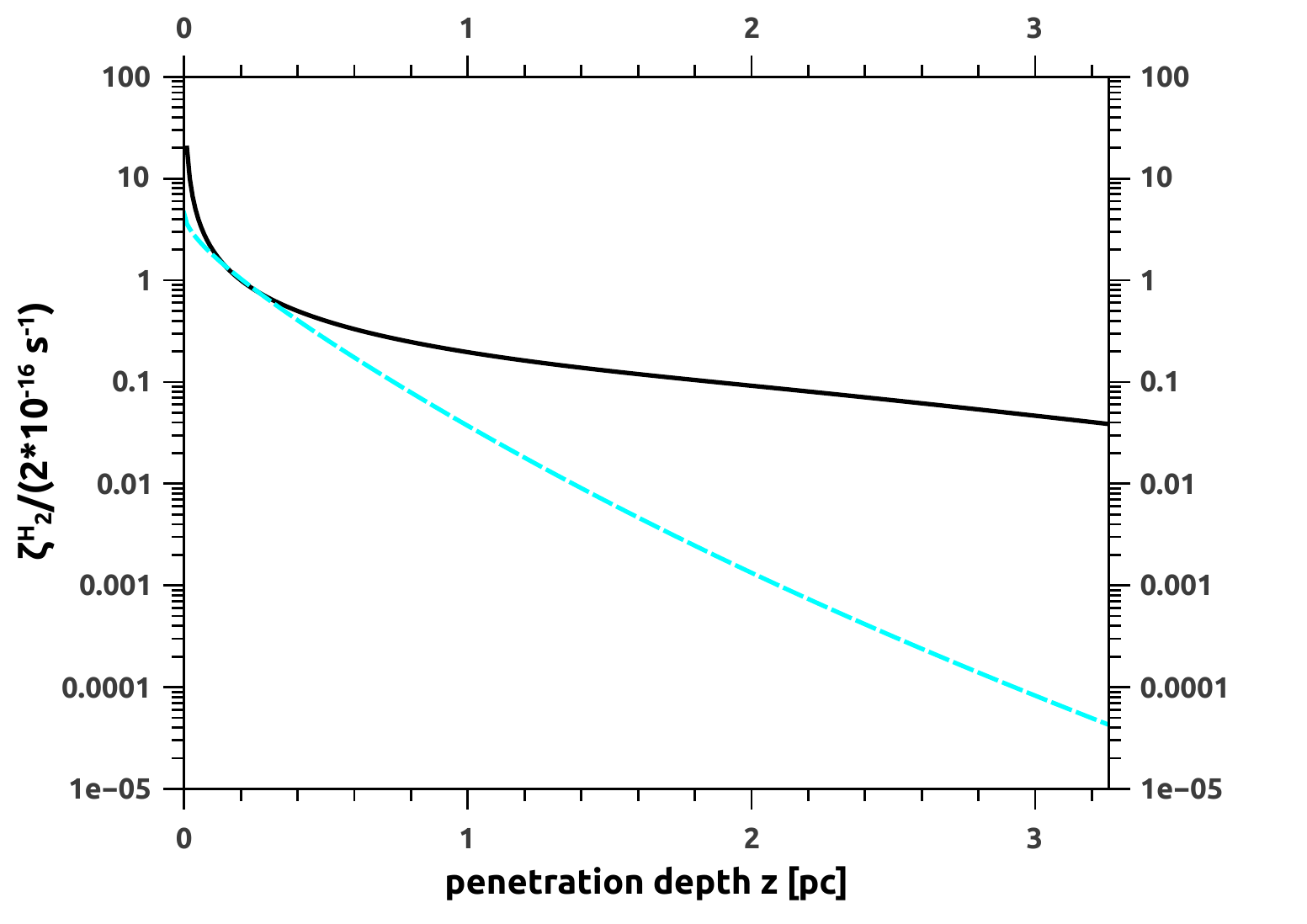}}%
\caption[...]
 {CR-induced ionization profiles for the objects W49B \subref{W49B_dyn_diff}, W44 \subref{W44_dyn_diff}, 
 3C~391 \subref{3C391_dyn_diff}, and CTB~37A \subref{CTB37A_dyn_diff}, originating from diffusion-driven motion 
 (black, solid lines) and relativistic, kinematic motion (cyan, long-dashed lines), respectively.}%
\label{ion_dyn_diff}%
\end{figure}

%================================================
The CR- and photoinduced ionization profiles for W49B, W44, 3C~391, and CTB~37A computed in Sects.\ \ref{CR_ion_prof} and \ref{X_att}, 
respectively, are shown and compared. 
The cases of purely diffusion-driven motion as well as purely relativistic, kinematic motion of the CR protons and 
their implications for the CR-induced ionization rate are shown in Fig.\ \ref{ion_dyn_diff}. 

\noindent For the specific diffusion coefficient assumed here, the diffusion-driven motion dominates the relativistic, kinematic 
motion for all considered particle momenta. 
This can be seen from the lower overall magnitude of the ionization rate for kinematic motion, indicating that the 
kinematically driven particles take longer to reach a certain penetration depth and consequently lose more energy 
at low penetration depths. This in turn means that fewer particles are capable of ionizing the molecular hydrogen 
at large penetration depths. 
This relation between the diffusion and the relativistic speeds, $v_{\mathrm{diff}}>v_{\mathrm{rel}}$, 
arises because the magnetic field inhomogeneities that determine the 
diffusion coefficient effectively boost the CR particles to diffusion speeds exceeding their relativistic speeds. 
Since the diffusion-driven and the relativistic, kinematic motions occur simultaneously, the effective 
particle speed is a combination of both. However, because the former is dominant, it is used as an 
approximation for the effective CR particle speed. 

\begin{figure}[ht]%
\centering
\subfigure[][]{%
\label{W49B_diff_X}%
\includegraphics[width=0.48\columnwidth]{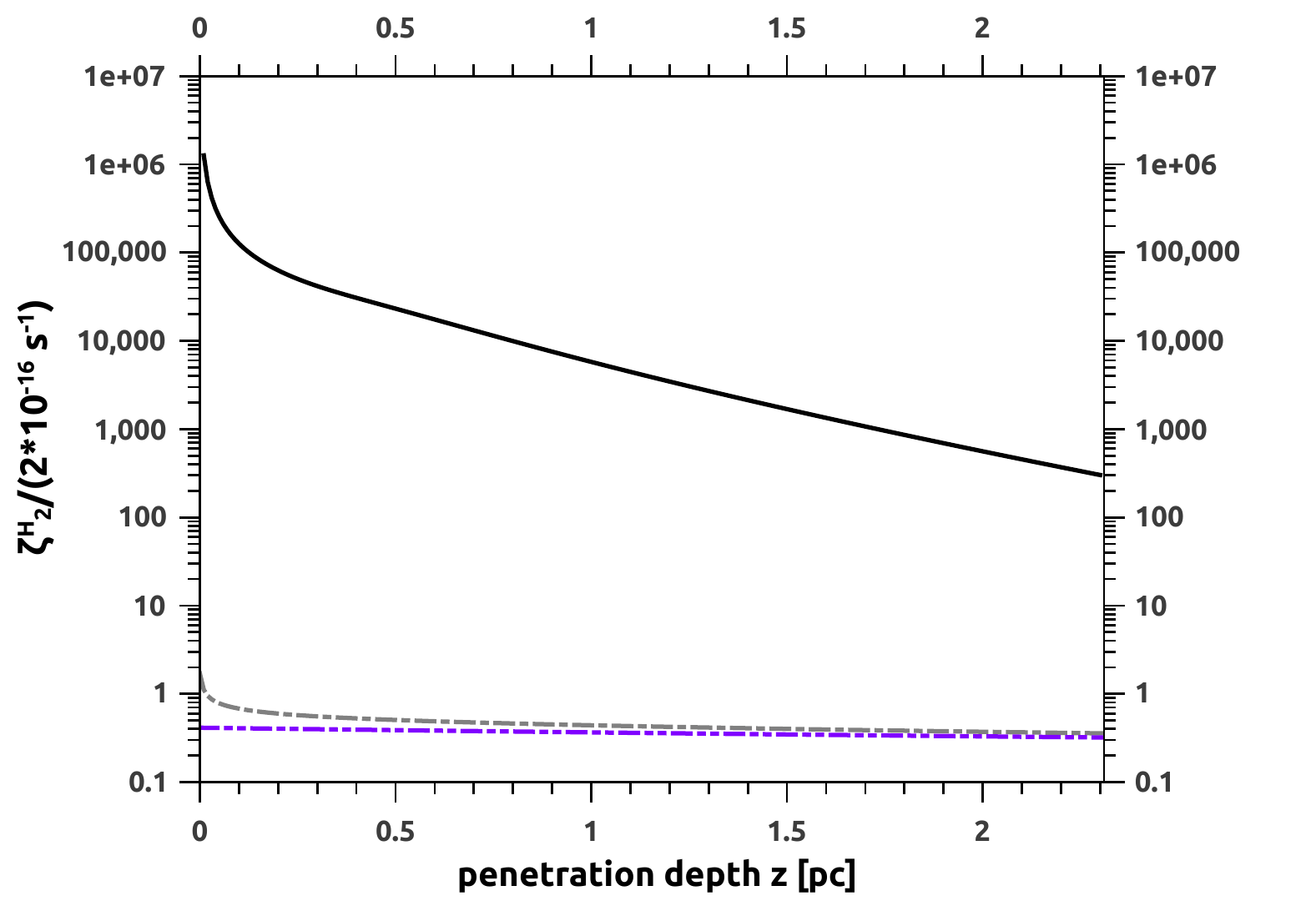}}%
\hspace{8pt}%
\subfigure[][]{%
\label{W44_diff_X}%
\includegraphics[width=0.48\columnwidth]{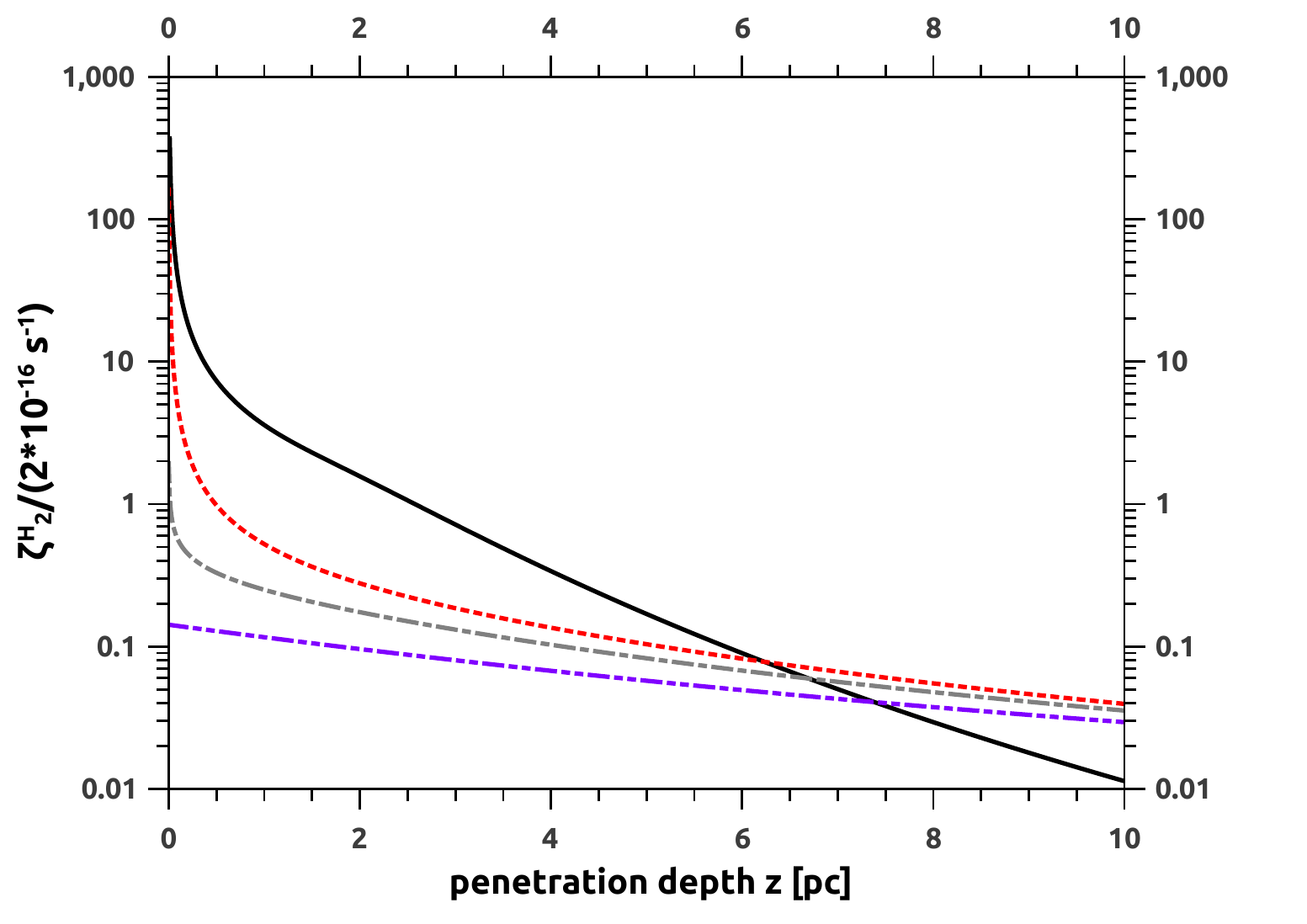}}\\
\subfigure[][]{%
\label{3C391_diff_X}%
\includegraphics[width=0.48\columnwidth]{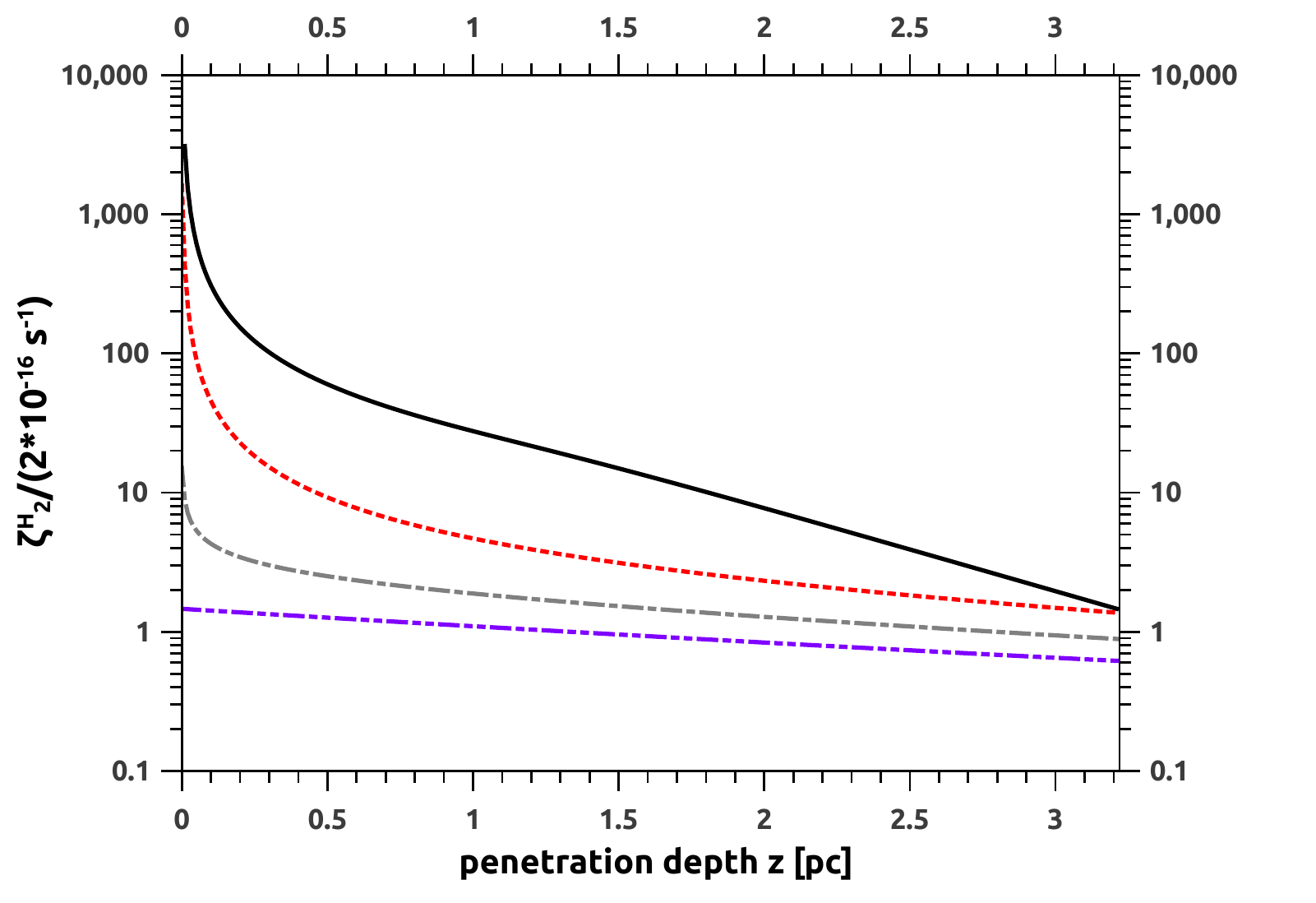}}%
\hspace{8pt}%
\subfigure[][]{%
\label{CTB37A_diff_X}%
\includegraphics[width=0.48\columnwidth]{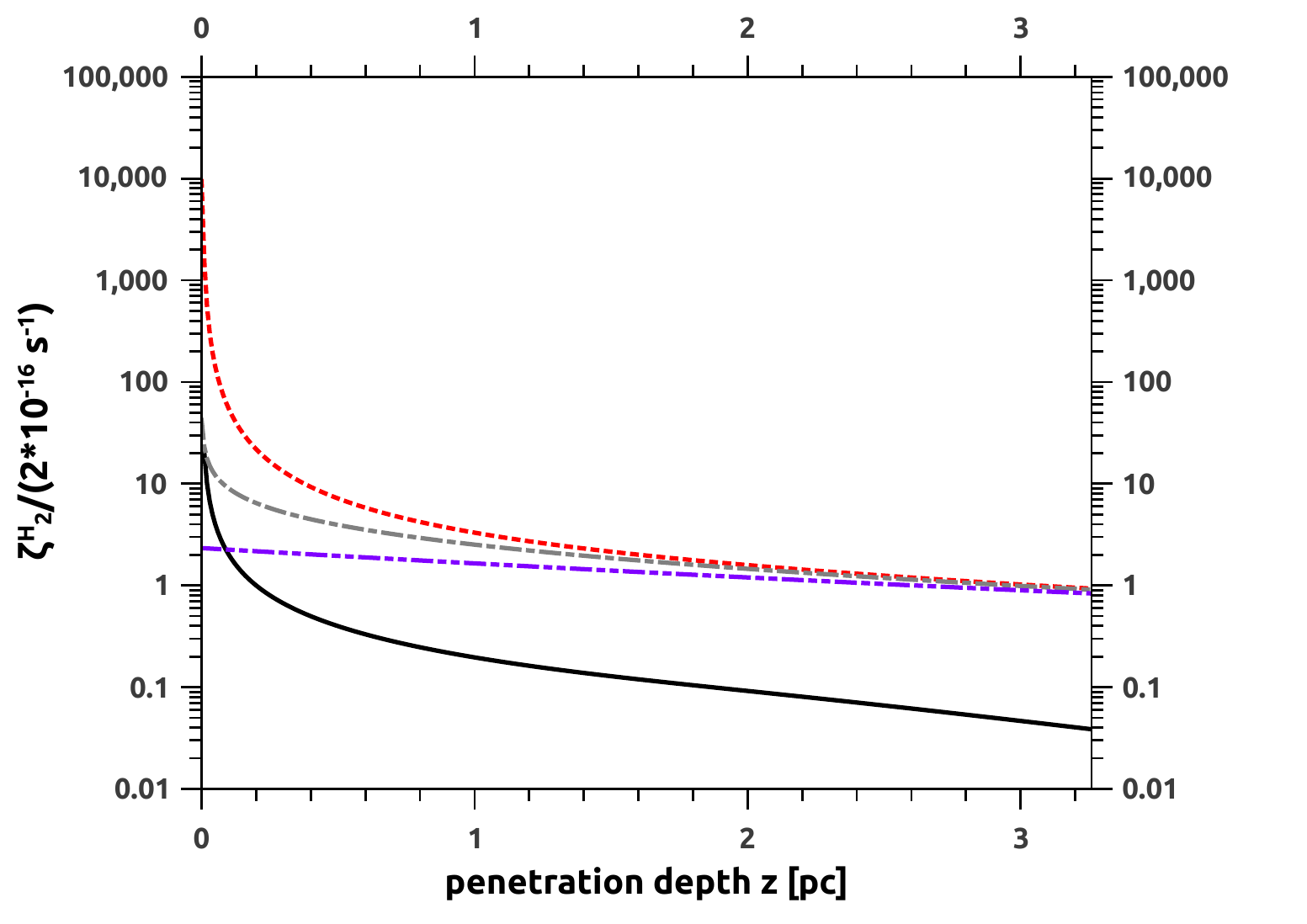}}%
\caption[...]
 {Ionization profiles for the objects W49B \subref{W49B_diff_X}, W44 \subref{W44_diff_X}, 3C~391 \subref{3C391_diff_X}, 
 and CTB~37A \subref{CTB37A_diff_X}, induced by cosmic rays (black, solid lines), as well as X-ray spectra extrapolated 
 toward lower energies by a power-law (red, short-dashed lines), by a constant (gray, long-short-dashed lines), and 
 without extrapolation (purple, long-short-short-dashed lines), respectively.}%
\label{ion_diff_X}%
\end{figure}

In Fig.\ \ref{ion_diff_X}, the photoionization profiles for all three low-energy extrapolations of the 
X-ray data are presented, and compared with the ionization profiles induced by CRs. 
The power-law extrapolation (I) toward the UV photon energy domain leads to overestimates of the photon fluxes and, 
therefore, to reasonable upper limits of the photoinduced ionization rates, while the cut-off X-ray fluxes (III) yield 
lower limits of the photoionization rates. 
Particularly at very short penetration depths, the ionization rates differ distinctly, because on the one hand, 
at low photon energies the total photoabsorption cross-section is large, which means that low-energy photons are absorbed very efficiently 
and, on the other hand, the parametric descriptions of the photon fluxes at low energies are different. 
With increasing penetration depth, the different photoionization profiles become asymptotically equivalent, which 
has two reasons. 
On the one hand, at high photon energies the total photoabsorption cross-section is small, and 
on the other hand, the parametric descriptions of the photon fluxes at high energies are identical. 

\begin{figure}[ht]%
\centering
\subfigure[][]{%
\label{W49B_tot_X}%
\includegraphics[width=0.48\columnwidth]{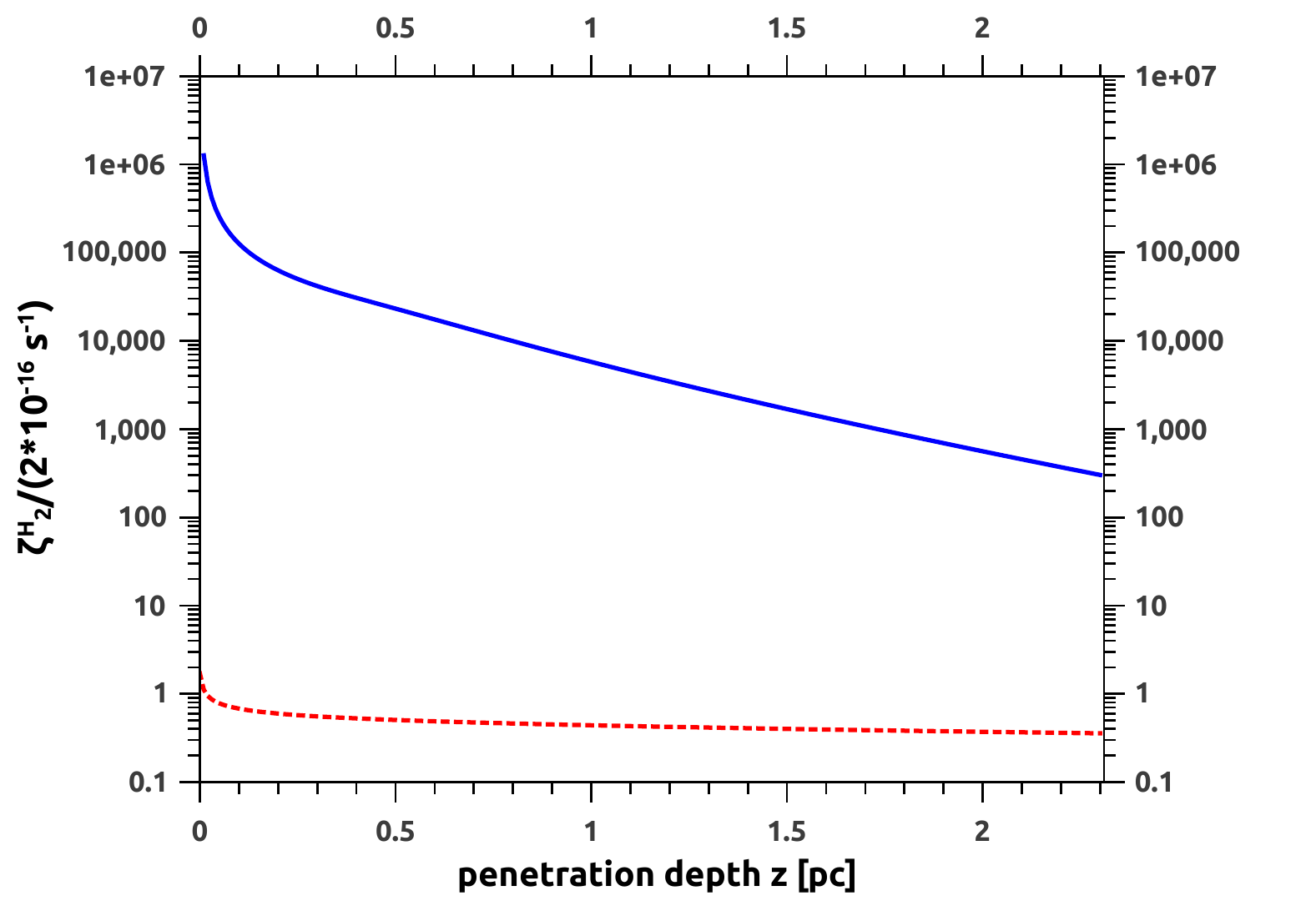}}%
\hspace{8pt}%
\subfigure[][]{%
\label{W44_tot_X}%
\includegraphics[width=0.48\columnwidth]{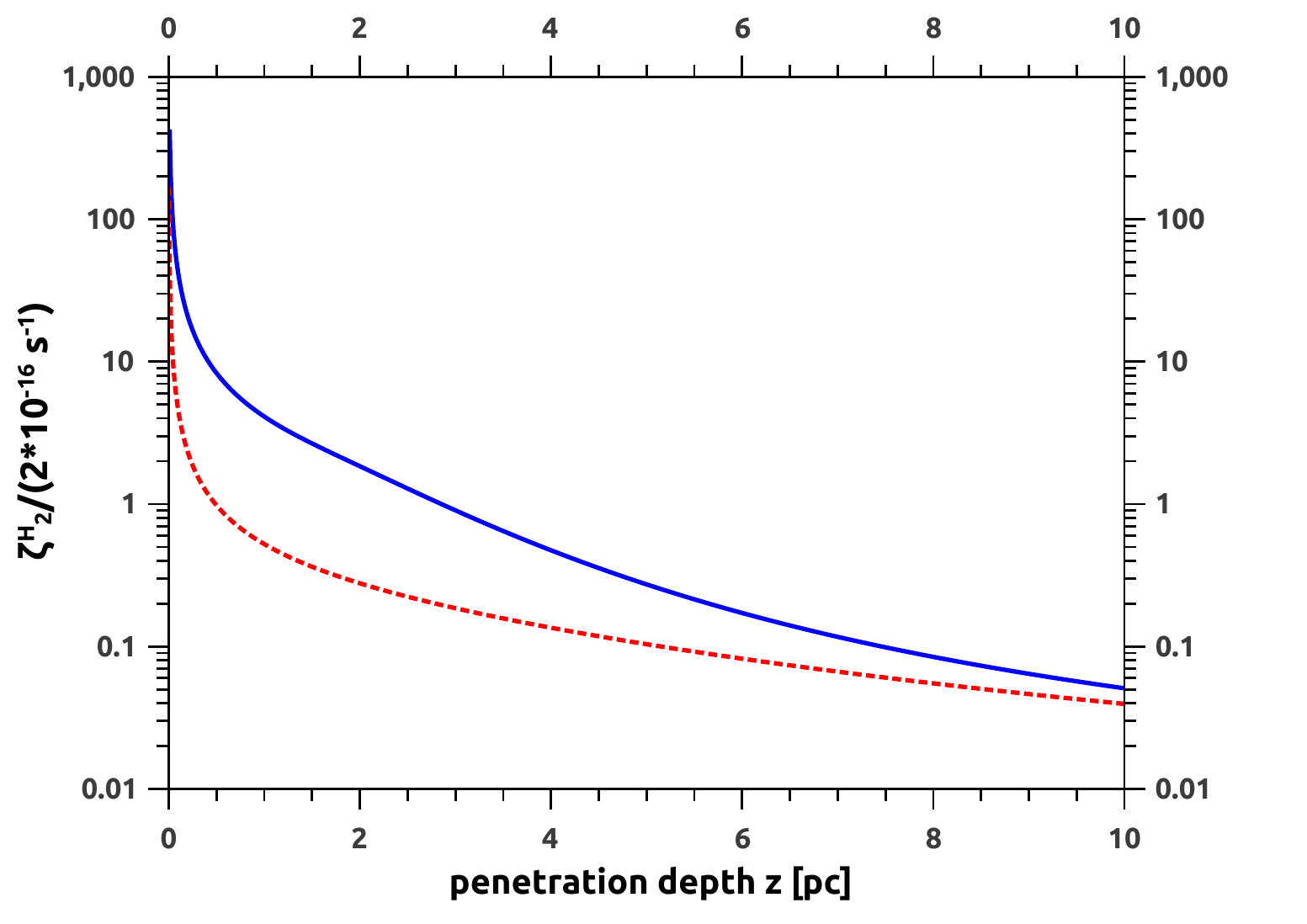}}\\
\subfigure[][]{%
\label{3C391_tot_X}%
\includegraphics[width=0.48\columnwidth]{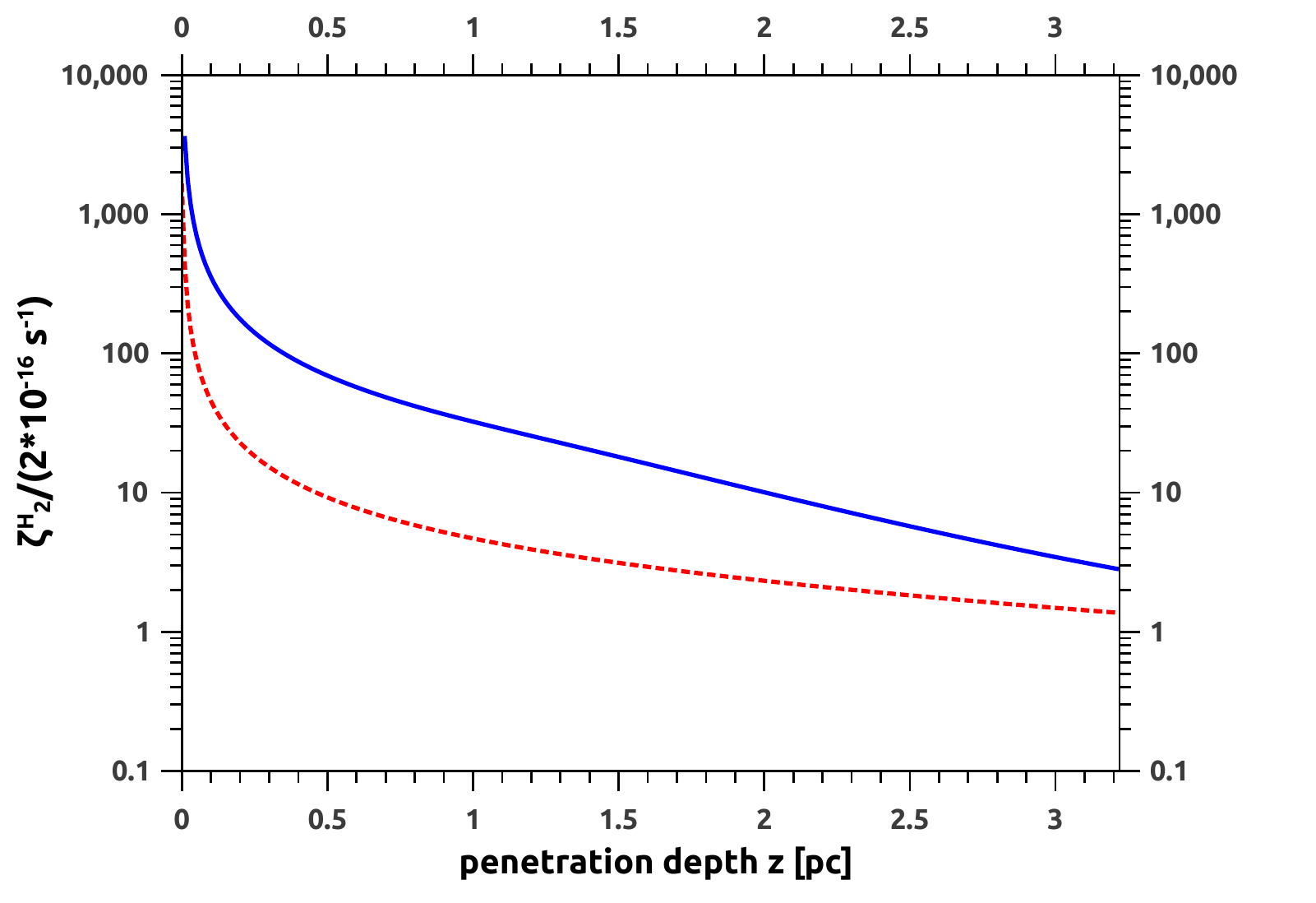}}%
\hspace{8pt}%
\subfigure[][]{%
\label{CTB37A_tot_X}%
\includegraphics[width=0.48\columnwidth]{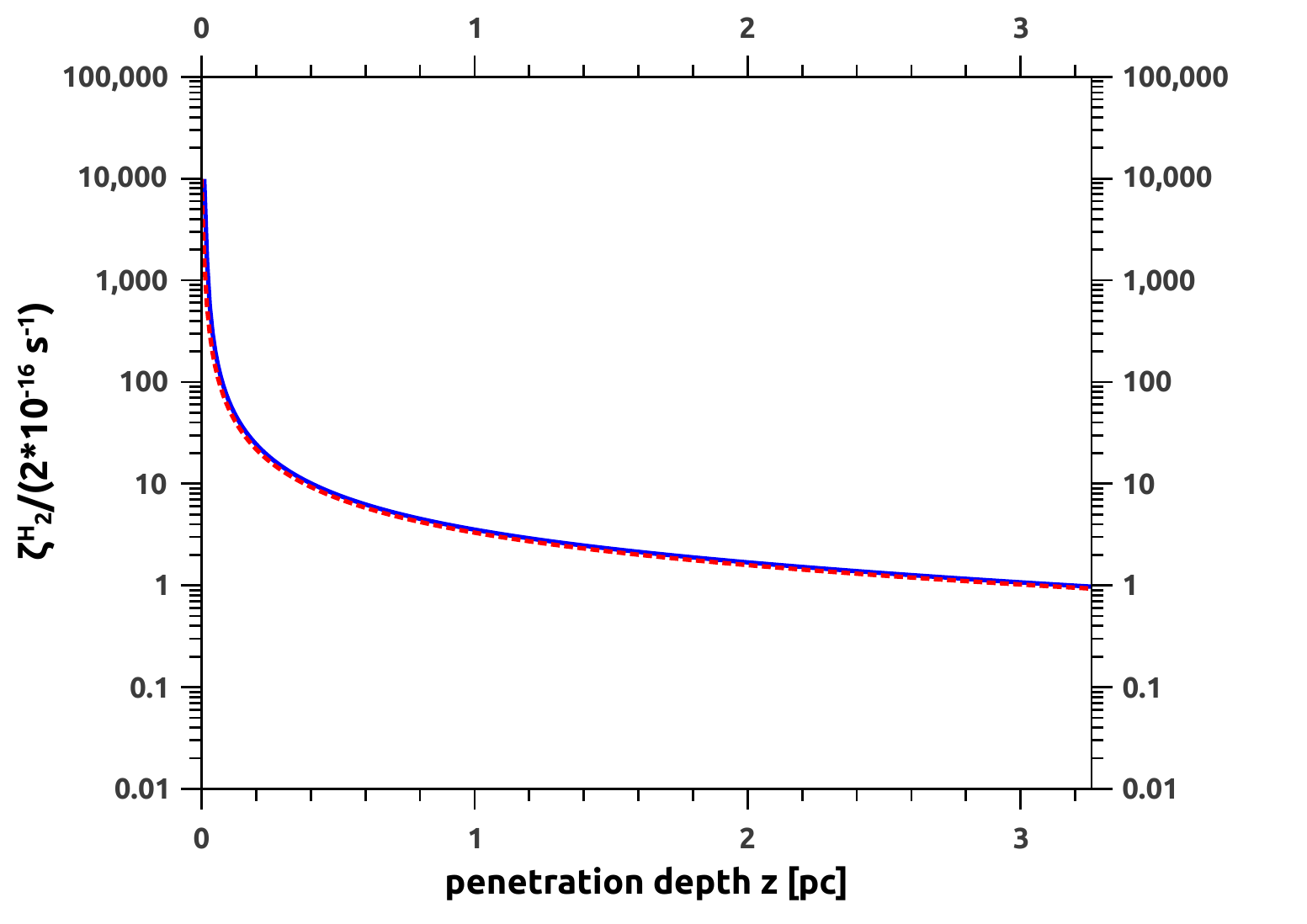}}%
\caption[...]
 {Total ionization profiles (blue, solid lines) and ionization profiles induced by X-ray spectra extrapolated 
 toward lower energies by a power-law alone (red, dashed lines), respectively, for W49B \subref{W49B_tot_X}, 
 W44 \subref{W44_tot_X}, 3C~391 \subref{3C391_tot_X}, and CTB~37A  \subref{CTB37A_tot_X}.}%
\label{ion_tot_X}%
\end{figure}

Among the four objects studied in this work, W49B shows the highest CR-induced ionization rate, 
more than two orders of magnitude higher than the corresponding ionization rate found for 3C~391, 
which has the second-highest CR-induced ionization rate of the objects studied here. %\\
The rather flat photoinduced ionization profile for W49B results from the flat spectral shape of 
the photon flux (cf.\ Fig.\ \ref{fits_X}). 
The ionization in this cloud is clearly dominated by CRs. In CTB~37A, the 
influence of photons on the ionization significantly exceeds the influence of 
CR protons. Figure \ref{CTB37A_diff_X} indicates that this behavior does not depend on the extrapolation of the 
X-ray flux toward lower photon energies, because even without any extrapolation (III), the photoinduced 
ionization rate dominates, except for the very surface region of the MC. 
For both W44 and 3C~391, the CR-induced ionization dominates up to a certain penetration depth, $6.5\,$pc 
for W44 and $3.2\,$pc for 3C~391, while at larger penetration depths photons dominate the 
ionization. 
Since the parametric descriptions (I) of the low-energy photon fluxes used for these calculations are overestimates of 
the real photon fluxes, the actual photoionization rates are lower. 
Moreover, it should be kept in mind that the effects of magnetic fields, that is, magnetic focusing and magnetic mirroring, 
on the CR-induced ionization rates have not been taken 
into account in this approach. \cite{padovani2011} considered these effects in their studies, and came 
to the conclusion that the influence of magnetic fields, on average, decreases the ionization rate induced by 
CR protons by a factor of $3$ to $4$. For the rather diffuse clouds examined here, the decrease 
of the CR-induced ionization rate would drop to a factor of $3$ or less. This would not change the overall trend of which 
ionization process is dominant in the studied SNR-MC systems, because the ionization profiles 
induced by CR protons and photons differ by at least a factor of 10 at all relevant penetration depths for both W49B 
and CTB~37A. For W44 and 3C~391, 
they differ by approximately a factor of 10 for a wide range of penetration 
depths even if extrapolation (I) is used. 
For W44, 3C~391, and CTB~37A, volume-filling factors of $1$ are applied. While this assumption is common, 
a volume-filling factor $\ll1$ is more realistic. Taking this into consideration, the CR proton flux normalizations 
would increase, and so would the CR-induced ionization rates. 
Hence, for 
W44 and 3C~391, an observational distinction between cosmic rays and photons as the dominant source of ionization seems feasible. 
For the calculation of the CR proton normalization for W49B, a volume-filling factor of $0.06$ was applied 
\citep{abdo(W49B)2010}. Changing this factor to $1$ would yield a CR-induced ionization rate that would still be the highest 
among all the objects considered and would also be orders of magnitude higher than the corresponding ionization rate induced by 
photons. %\\ 
It should be pointed out that the total ionization rates very close to the surfaces of the MCs, i.e., 
in the outer $\sim0.1\,\mathrm{pc}$, strongly depend on the estimates used to model the EUV regime. 
In these regions, the contribution of EUV radiation to the total ionization rate is highest 
(cf.\ the rapid decline of the photoinduced ionization rate at very low 
penetration depths in Fig.\ \ref{ion_diff_X}, particularly 
the difference between the different extrapolations of the X-ray fluxes seen in Figs.\ \ref{W44_diff_X}, 
\ref{3C391_diff_X}, and \ref{CTB37A_diff_X}). %\\
Since the ionization profiles induced by different processes cannot be detected separately, 
it is reasonable to compare the total ionization rates with the ionization rates expected from photoinduced ionization alone. 
Detecting an ionization rate that clearly exceeds the theoretical photoinduced ionization rate (enhanced ionization rate) 
would imply a significant contribution from low-energy CR protons, indicating 
that high-energy CR protons are the driving force for gamma-ray emission. 
In the following, for a conservative comparison, only the extrapolation of the X-ray fluxes toward lower photon energies 
by the power-law (I) is discussed. 
The total ionization profiles analyzed here are simply the sum of the photoinduced ionization profiles, with 
the power-law extrapolation (I), and the CR-induced ionization profiles, with $v=v_{\mathrm{diff}}$ as the 
effective CR particle speed and the corresponding diffusion time $t=T_{\mathrm{diff}}$. 
A comparison of the total ionization profiles and the photoinduced ionization profiles for each object is 
shown in Fig.\ \ref{ion_tot_X}. 

\noindent A huge difference between the CR- and photoinduced ionization profiles implies that the total 
ionization rate almost coincides with the ionization profile from the dominant process in the MC. 
For W49B, the total ionization rate practically coincides with the CR-induced ionization rate (cf.\ Fig.\ \ref{W49B_diff_X}), 
while for CTB~37A, it coincides with the ionization rate induced by photons (cf.\ Fig.\ \ref{CTB37A_diff_X}). 
The impact of the calculated ionization rates on the chemistry inside the MCs, with a focus on the detectability 
of ionization rates, is discussed in the following section.

%================================================
\section{Conclusions and outlook\label{conclusions}}
%================================================
We calculated the position-dependent H$_2$ ionization rates of molecular clouds near supernova remnants, induced by photons 
and cosmic-ray protons, respectively, and compared them to examine a potential hadronic formation scenario 
of the observed gamma radiation from these systems. 
The cosmic-ray proton fluxes were constructed by means of the analytic solution of the transport equation for cosmic-ray 
protons inside the clouds, while the photon fluxes were obtained using the Beer-Lambert law. 
In addition to the systems consisting of a supernova remnant and a molecular cloud that we studied here, 
the solution of the transport equation can also be used for many other astrophysical objects with scalar, 
momentum-dependent diffusion and any type of momentum losses, such as 
stellar winds, cosmic-ray diffusion in the interstellar medium, or gamma-ray bursts. 

Enhanced ionization rates compared with the ionization rates induced by photons alone are expected if the 
gamma-ray emission 
seen from systems containing supernova remnants and molecular clouds originates from the decay of neutral pions that are formed 
in proton-proton interactions of the cosmic-ray protons with ambient hydrogen. The method presented here is particularly 
promising for objects of low X-ray fluxes and steep, incident cosmic-ray proton spectra, when the spectra 
can be extrapolated toward lower energies by single power-laws. %\\
For the SNR-MC system W49B, the ionization rate induced by cosmic-ray protons significantly exceeds the ionization rate induced by photons. 
In contrast, for CTB~37A, the total ionization rate is dominated by photoinduced ionization, while for the objects W44 and 3C~391, 
the results presented here favor cosmic-ray protons as the dominant source of ionization. 
The results shown in Figs.\ \ref{ion_diff_X} and \ref{ion_tot_X} are a definite motivation for further studies 
of enhanced ionization rates attributable to cosmic-ray sources. 
But since the ionization rate is not directly detectable, one has to derive it for instance from observational spectroscopic data. 
One method to estimate the ionization rate is to examine the absolute intensity of molecular rotation-vibrational lines, as stated 
in \cite{becker2011}. It is also possible to analyze the ratio of the abundance of certain molecules 
or molecular ions \citep[see, e.g.,][]{odonnell1974, black1977, indriolo2009}. 
Computing rotation-vibrational lines requires solid estimates on several physical parameters of the molecular clouds, such 
as composition, temperature, ionization degree, electron density, and a few others. 
Many of these parameters are quite difficult to estimate, which means that predictions for rotation-vibrational line 
intensities are inaccurate. Hence, the molecular ion H$_3^+$ is especially well-suited for observational examinations of 
ionization rates because only very few cloud parameters are involved in calculating its formation rate \citep{goto2013}. 
Particularly in W49B and 3C~391, very high ionization rates are found ($10^{-12}\,$s$^{-1}$ at depths 
of 1\,pc into clouds with a hydrogen density of n$_{\rm H}=100\,$cm$^{-3}$). The resulting ionization degrees are on the order of $10^{-4}$ at 
depths of 1\,pc and $10^{-3}$ or higher at the cloud surfaces. However, a penetration depth of 1\,pc at a density of 
100\,cm$^{-3}$ corresponds to a column density of $3\cdot10^{20}$\,cm$^{-2}$, so these regions will also be significantly 
affected by stellar Far UV radiation ($E<13.6\,$eV), which has an impact on the chemistry and temperature in the clouds. 
Accordingly, it is also possible that these outer layers of the clouds are photodissociation regions. This needs 
to be taken into account in the astrochemistry models for the prediction of molecular rotation-vibrational 
lines to test the theoretical ionization rates by means of observations.

\begin{table}[htb]
\centering{
\begin{tabular}{ccccccc}
	\hline\hline
  instrument        &    H.E.S.S.    & \multicolumn{2}{c}{FermiLAT}  & CTA & \multicolumn{2}{c}{ALMA} \\
  \hline  
  energy or $\lambda$ &    >100 GeV    &    0.1 GeV     &    100 GeV     &  10 TeV  &   72.5\,mm    &    362.5\,$\mu$m    \\
  %\hline  
  ang.\ resolution & 0.1$^{\circ}$  &  3$^{\circ}$   & 0.04$^{\circ}$ &  1' &     1''    &    0.005''    \\
  %\hline  
  object  &        \multicolumn{6}{c}{linear spatial resolution for 90$^{\circ}$ inclination [pc]}   \\
  \hline
  W49B    &             14.0      &      419       &      5.6       & 2.3 &   0.0388   &    0.000194   \\
  W44     &              5.8      &      173       &      2.3       & 1.0 &   0.0160   &    0.000080   \\
  CTB 37A &             14.0      &      419       &      5.6       & 2.3 &   0.0388   &    0.000194   \\
  3C 391  &             12.6      &      377       &      5.0       & 2.1 &   0.0349   &    0.000175   \\
  \hline
\end{tabular}
\vspace{0.5cm}
\caption{Spatial resolutions of the studied SNR-MC systems for a selection of relevant telescopes. The distances of the 
objects from Earth were taken from Table \ref{tab_dist_ext}. The corresponding relative errors also 
apply to the spatial resolutions presented here. For ALMA, the maximum base length of 14.5\,km was used. 
Sources for the instrumental data: H.E.S.S.: \cite{HESSresolution}, FermiLAT: {http://www-glast.stanford.edu/instrument.html}, 
CTA: {http://www.cta-observatory.org/?q=node/26}, ALMA: \cite{ALMAresolution}.}
\label{resolution}}
\end{table}

\noindent Some of the molecular species linked to high ionization rates are most effectively and hence solely 
detectable in absorption. This means that the molecular clouds in which these molecules are to be found have to be somewhat close to Earth, and 
the sightlines toward these clouds need to feature a bright source in the background for illumination such that 
the light is absorbed by the molecules in order to trace them. This might restrain the target-selection process of
systems composed of a supernova remnant associated with molecular clouds to nearby complexes, that is, a few kpc from Earth, 
such as W44. These are particularly interesting because the resolution available with current and next-generation telescopes will 
allow studying their substructure. 
The currently most promising experiments for the suggested correlation study are the Cherenkov Telescope Array (CTA) 
and FermiLAT for the detection of the gamma rays, and the Atacama Large Millimeter/submillimeter Array (ALMA) for the 
detection of the molecular lines. 
CTA will reach a spatial resolution of $\sim0.03^\circ$ \citep{cta2010,cta2011}. While this is still lower than for submm, 
IR, or FIR telescopes, CTA measurements can be used to determine regions of enhanced gamma-ray emission that then can 
be observed with IR telescopes to search for a correlation predicted for hadronic emission scenarios. 
The spatial resolution for a selection of relevant telescopes for the suggested correlation study is given in Table \ref{resolution}. 
Since different observations indicate high ionization rates through the measurement of H$_3^+$ \citep[see, e.g.,][]{indriolo2010}, 
it is possible to find the ionization-induced signatures that are spatially correlated with the GeV gamma-ray emission, which 
will help pinpointing supernova remnants as sources of cosmic-ray protons. 

Apart from the specific supernova remnants discussed above, there are currently six additional ones that are associated 
with molecular clouds. However, for most of these no suitable X-ray data are currently available, because the X-ray fluxes 
in the corresponding energy ranges are either below the detection limit or are not aligned with the gamma-ray emission regions. 
For these systems, precise measurements of X-ray spectra by experiments such as SWIFT, XMM-Newton, Suzaku, or Chandra, particularly 
for energies $E\sim0.1\,$keV, can help to estimate the photoionization rate. 

To obtain even more accurate theoretical predictions for ionization profiles induced by cosmic rays, 
additional components of both the cosmic-ray spectrum and the cloud matter can be taken into account. Considering 
also heavier CR particles in addition to protons for calculating the ionization rate of molecular hydrogen, the model can 
be applied by adequately adapting the Coulomb-loss coefficient and the ionization cross-section, for example by using 
the Bethe-Bloch approximation \citep{bethe1933,padovani2009}. 
The relative abundances of cosmic-ray helium or heavier nuclei at the acceleration regions are poorly known, however.  
In addition to the ionization of molecular hydrogen in the clouds, the model can also be used to calculate the 
ionization rates of other species such as helium by simply replacing the ionization cross-section. This 
further improves estimates of the ionization degree in the molecular clouds required for the chemistry models. %\\

To conclude, observable ionization signatures are expected and, combined with gamma-ray observations, correlations 
that can only be caused by a common primary component \citep[see, e.g.,][]{juliareview2008} can be revealed.

%================================================
\section*{Acknowledgments\label{ack}}
%================================================
We would like to thank R.\ Schlickeiser, J.~H.\ Black, and S.\ Sch\"oneberg for helpful and inspiring discussions. 
We acknowledge helpful comments and suggestions from the anonymous referee, which considerably improved 
the quality of the manuscript. 
We are also grateful to M.\ Ozawa for providing the X-ray data from Chandra for W49B. Furthermore, we thank 
Hongquan Su for informing us about revised observational values for the distance of W44, CTB~37A, and 3C~391, and 
H.\ Fichtner for a careful reading of the manuscript. 
FS also thanks M.\ Mandelartz for helpful advice concerning programming. 

\bibliographystyle{aa}
\bibliography{lib_a_and_a.bib}

\newpage

%================================================
\section*{Appendix A: General solution of the transport equation\label{app_A}}
%================================================
In order to find a general solution of the transport equation (\ref{transp_eq_large}) using Green's method, first the 
fundamental solution $G(\vec{r},p,t\,|\,\vec{r}_0,p_0,t_0)$, that is, the solution for the Dirac source distribution 
$$Q(\vec{r},p,t) = Q_0(\vec{r},p,t) = \delta^3(\vec{r} - \vec{r}_0)\delta(p - p_0)\delta(t - t_0),$$ is determined. 
The fundamental transport equation is then given by
\begin{equation}
\frac{\partial G}{\partial t} - D(p) \Delta G 
- \frac{\partial}{\partial p}\big(b(p) \cdot G \big) 
= \delta^3(\vec{r} - \vec{r}_0)\delta(p - p_0)\delta(t - t_0).
\label{transp_eq}
\end{equation}
In terms of the function $R(\vec{r},p,t) = b(p) \cdot G(\vec{r},p,t)$\label{def_R}, 
Eq.\ (\ref{transp_eq}) becomes
\begin{equation}
\frac{\partial R}{\partial t} - D(p) \Delta R - b(p) \frac{\partial R}{\partial p} 
= b(p)\;\delta^3(\vec{r} - \vec{r}_0)\delta(p - p_0)\delta(t - t_0).
\label{eq:diff_R_p}
\end{equation}
With the new coordinate 
\begin{equation}
\beta - \beta_0 = -\left.\frac{m_{\mathrm p}c}{3\,a_{\mathrm{ad}}}\ln \left(a_{\mathrm{cc}} 
+ a_{\mathrm{ad}} \left(\frac{p}{m_{\mathrm p}c} \right)^3 \right) 
 \right|_{p_0}^{p} \label{eq_help_beta}, 
\end{equation}
induced by $\partial_\beta=-b(p)\partial_p$, and 
$\delta(\beta - \beta_0) = b(p_0) \cdot \delta(p - p_0)$, 
Eq.\ (\ref{eq:diff_R_p}) yields
\begin{equation}
\frac{\partial R}{\partial t} - D(\beta) \Delta R + \frac{\partial R}{\partial \beta} 
= \delta^3(\vec{r} - \vec{r}_0)\delta(\beta - \beta_0)\delta(t - t_0).
\label{eq:diff_R}
\end{equation}
This partial differential equation can be solved by first applying a Laplace transformation 
$\mathfrak{L}[\,\cdot\,]$ with respect to the time variable $t$ 
\begin{equation}
\mathfrak{L}[R] \equiv \mathfrak{G}(\vec{r},\beta,s) = \int_0^\infty {\exp(-st) R(\vec{r},\beta,t)\,\mathrm d}t. \label{Lapl_t}
\end{equation}
Note that the lower integration limit in the definition of the Laplace transformation (\ref{Lapl_t}) 
fixes $t_0=0$ without loss of generality. 
The Laplace transform of Eq.\ (\ref{eq:diff_R}) reads
\begin{align}
-R(\vec{r},\beta,t=0) + s\,\mathfrak{G}(\vec{r},\beta,s) - D(\beta) \Delta \mathfrak{G}(\vec{r},\beta,s) 
+ \partial_\beta \mathfrak{G}(\vec{r},\beta,s) %\nonumber \\ 
= \delta^3(\vec{r} - \vec{r}_0)\delta(\beta - \beta_0).
\label{Lapl_R}
\end{align}
Since $R$ is related linearly to the differential CR proton number density, $R\propto \mathrm{d}N/\mathrm{d}V,$ where $N$ is the (finite) 
number of protons, and $V$ is the volume within which the protons are distributed, this function vanishes 
at $t=0$, because the distribution of CR protons at this time is restricted to the boundary $\vec{r}=\vec{r}_0$ of 
the MC.
Then, Eq.\ (\ref{Lapl_R}) reduces to the inhomogeneous, linear partial differential equation of first 
order in $\beta$ and second order in $\vec{r}$, 
\begin{equation}
\Delta \mathfrak{G} - \frac{1}{D(\beta)}[s + \partial_{\beta}] \mathfrak{G} 
= - \frac{\delta(\beta - \beta_0)}{D(\beta)}\delta^3(\vec{r} - \vec{r}_0).
\label{eq:diff_G_L1}
\end{equation} 
This equation can be solved explicitly by using Duhamel's principle \citep{duhamel1838, duhamel}, 
which is a general method to find solutions of inhomogeneous, linear 
partial differential equations in terms of the solutions of the Cauchy problems for the corresponding homogeneous 
partial differential equations, that is, by interpreting the inhomogeneity as a boundary value condition in a 
higher-dimensional space labeled with an additional auxiliary variable. 
This principle is applied here on a linear partial differential equation with a product-separable inhomogeneity of 
single-variable factors with respect to the variables $\vec{r}$ and $\beta$, 
\begin{equation}
 \big(\hat{O}_{\vec{r}} + \hat{O}_{\beta,s}\big)\mathfrak{G}(\vec{r},\beta,s)
 = -F_1(\vec{r})F_2(\beta),
\label{prod-sep}
\end{equation}
where 
\begin{equation}
\hat{O}_{\vec{r}} = \Delta, \,\,\, \hat{O}_{\beta,s} = -\frac{1}{D(\beta)}\left[s+\partial_\beta\right], 
\,\,\, F_1(\vec{r}) = \delta^3(\vec{r}-\vec{r}_0), 
\,\,\, \mathrm{and}\,\,\, F_2(\beta)= \frac{\delta(\beta - \beta_0)}{D(\beta)}.
\end{equation}
From the structure of the left-hand side of Eq.\ (\ref{prod-sep}), it directly follows that the solution of the 
corresponding homogeneous equation is product-separable $\mathfrak{G}_{\mathrm{hom}}=S(\vec{r})P(\beta,s)$. 
An embedding of the original $(\vec{r},\beta,s)$-space into the extended, higher-dimensional $(\vec{r},\beta,s,u)$-space, 
where $u\in\mathbb{R}^+_{0}$, implies that there is a family of functions \mbox{$S_u(\vec{r}):=S(\vec{r},u)$} and 
$P_u(\beta,s):=P(\beta,s,u)$, fulfilling the relations $\hat{O}_{\vec{r}}S(\vec{r},u) = \partial_u S(\vec{r},u)$ and 
$\hat{O}_{\beta,s}P(\beta,s,u) = \partial_u P(\beta,s,u)$ with the boundary conditions 
\begin{equation}
S(\vec{r},u=0)=F_1(\vec{r}),\,\,P(\beta,s,u=0)=F_2(\beta),\,\,\mathrm{and}\,\,S(\vec{r},u=\infty)=P(\beta,s,u=\infty)=0,
\label{eq:schlicky_trick_bc}
\end{equation}
such that the full, non-trivial solution of Eq.\ (\ref{prod-sep}) is given by the integral 
\begin{equation}
\mathfrak{G}(\vec{r},\beta,s) = \int_0^\infty{S(\vec{r},u)P(\beta,s,u)\,\mathrm{d}u}.
\label{eq:schlicky_trick}
\end{equation}
Technically, this integral represents the "summation" over the entire family of homogeneous solutions in the extended coordinate 
space with boundary conditions compatible with the inhomogeneity. Within this setting, 
\mbox{Eq.\ (\ref{eq:diff_G_L1})} decouples into the simpler set of partial differential equations 
\begin{equation}
\frac{\partial S(\vec{r},u)}{\partial u} = \Delta S(\vec{r},u)
\label{eq:bc_1}
\end{equation}
and 
\begin{equation}
\frac{\partial P(\beta,s,u)}{\partial u} + \frac{1}{D(\beta)}\left[s+\partial_\beta\right]P(\beta,s,u) = 0,
\label{eq:bc_2}
\end{equation}
which has to be solved in order to determine the fundamental solution $\mathfrak{G}$ via the integral in 
Eq.\ (\ref{eq:schlicky_trick}). The solution of Eq.\ (\ref{eq:bc_1}) is the well-known heat kernel of 
three-dimensional Euclidean space
\begin{equation}
S(\vec{r},u) = \frac{1}{(4 \pi u)^{3/2}} \exp \bigg(- \frac{(\vec{r} - \vec{r}_0)^2}{4u} \bigg).
\label{eq:bc_1_sol}
\end{equation}
A solution of Eq.\ (\ref{eq:bc_2}) can directly be found after performing a Laplace transformation with 
respect to the \mbox{variable $u$}
\begin{equation}
\mathfrak{L}[P] \equiv \mathfrak{P}(\beta,s,q)=\int_0^\infty{\exp(-qu)P(\beta,s,u)\,\mathrm{d}u}.
\label{eq:Laplace_u-q}
\end{equation}
Then, Eq.\ (\ref{eq:bc_2}) becomes 
\begin{equation}
\big(q D(\beta) + s\big) \mathfrak{P} + \frac{\partial \mathfrak{P}}{\partial \beta} 
= \delta(\beta - \beta_0).
\label{eq:diff_frak_P}
\end{equation}
Note that the boundary condition $P(\beta,s,u=0)=\delta(\beta - \beta_0)/D(\beta)$ 
was already fixed in (\ref{eq:schlicky_trick_bc}). 
Using an integrating factor of the form 
$\exp\left(-s \beta - q \int_0^\beta{D(\beta')\,\mathrm{d}\beta'} \right)$, Eq.\ (\ref{eq:diff_frak_P}) 
can be rewritten as
\begin{eqnarray}
\frac{\partial}{\partial \beta} \left[\exp\bigg(s \beta + q \int_0^\beta{D(\beta')\,\mathrm{d}\beta'} \bigg) 
\cdot \mathfrak{P} \right] = \delta(\beta - \beta_0) \exp\bigg(s \beta + q \int_0^\beta{D(\beta')\,\mathrm{d}\beta'} \bigg)
\end{eqnarray}
and solved by simple integration, leading to 
\begin{equation}
\mathfrak{P}(\beta,s,q) = \Big(\Theta(\beta-\beta_0)+C(s,q)\Big)
\exp\bigg(-s(\beta-\beta_0)-q\int_{\beta_0}^{\beta}{D(\beta')\,\mathrm{d}\beta'}\bigg),
\end{equation}
where $\Theta(\cdot)$ is the Heaviside step function and $C(s,q)$ is an integration constant with respect to $\beta$.
Since only momentum-loss processes are considered, there are no particles with momenta larger than their initial momentum $p>p_0$, 
corresponding to $\beta < \beta_0$, at any time. Therefore, because the function $S(\vec{r},u)$ is independent of $\beta$, 
$\mathfrak{P}(\beta,s,q)$ must vanish for $\beta < \beta_0$ implying $C(s,q)=0$. 
Then, the solution of Eq.\ (\ref{eq:diff_frak_P}) reads 
\begin{equation}
\mathfrak{P}(\beta,s,q) = \Theta(\beta - \beta_0) 
\exp\bigg(-s(\beta - \beta_0) - q \int_{\beta_0}^\beta{D(\beta')\,{\mathrm d}\beta'}\bigg).
\label{eq:sol_frak_P}
\end{equation}
Via an inverse Laplace transformation with respect to the variable $q$, one can recover the function 
$P(\beta,s,u)$ 
\begin{equation}
P(\beta,s,u) = \mathfrak{L}^{-1}[\mathfrak{P}] = \frac{1}{2 \pi {\mathrm i}} 
\int_{c - {\mathrm i}\infty}^{c + {\mathrm i}\infty}{\exp(qu) \mathfrak{P}(\beta,s,q)\,\mathrm{d}q}.
\end{equation}
Since $\mathfrak{P}$ is well-defined and finite everywhere, one is free to choose the real-valued constant 
$c=0$. Therefore, from using the relation 
\begin{equation}
 \delta(x - x_0) = \frac{1}{2 \pi {\mathrm i}} 
\int_{-{\mathrm i}\infty}^{+{\mathrm i}\infty}{\exp\Big(w(x - x_0)\Big)\,\mathrm{d}w},
\end{equation}
it directly follows that
\begin{equation}
P(\beta,s,u) = \Theta(\beta - \beta_0) \exp\big(-s(\beta - \beta_0)\big) 
\delta\left(u - \int_{\beta_0}^\beta{D(\beta')\,{\mathrm d}\beta'}\right).
\label{eq:bc_2_sol}
\end{equation}
Substituting the functions (\ref{eq:bc_1_sol}) and (\ref{eq:bc_2_sol}) into the integral in Eq.\ (\ref{eq:schlicky_trick}) 
leads to 
\begin{align}
\mathfrak{G}(\vec{r}, \beta,s) = \frac{\Theta(\beta - \beta_0)\exp\big(-s(\beta - \beta_0)\big)}{(4 \pi)^{3/2}} %\nonumber \\ 
\cdot \int_0^\infty{u^{-3/2}\exp\left(- \frac{(\vec{r} - \vec{r}_0)^2}{4u}\right)
\delta\left(u - \int_{\beta_0}^\beta{D(\beta')\,{\mathrm d}\beta'}\right)\mathrm{d}u}. 
\end{align}
Because $ 0 \leq \int_{\beta_0}^\beta{D(\beta')\,{\mathrm d}\beta'} < \infty$, integration with 
respect to the variable $u$ yields
\begin{equation}
\mathfrak{G}(\vec{r}, \beta,s) = \frac{\Theta(\beta - \beta_0) \exp\big(-s(\beta - \beta_0)\big)}
{\left(4 \pi \int_{\beta_0}^\beta{D(\beta')\,\mathrm{d}\beta'}\right)^{3/2}} 
\exp\left(\frac{-(\vec{r} - \vec{r}_0)^2}{4 \int_{\beta_0}^\beta{D(\beta')\,\mathrm{d}\beta'}} \right).
\label{eq:frak_G_sol}
\end{equation}
In order to obtain $R(\vec{r},\beta,t)$ from $\mathfrak{G}(\vec{r},\beta,s)$, another inverse Laplace 
transformation, now with respect to the variable $s$, is performed, 
\begin{equation}
 \mathfrak{L}^{-1}[\mathfrak{G}] \equiv R(\vec{r},\beta,t) = \frac{1}{2 \pi {\mathrm{i}}}
\int_{c - {\mathrm{i}}\infty}^{c + {\mathrm{i}}\infty} {\exp(st) \mathfrak{G}(\vec{r},\beta,s)\,\mathrm d}s,
\label{eq:Laplace_t-s}
\end{equation}
where again $c=0$ is chosen, resulting in 
\begin{equation}
R(\vec{r},\beta,t) = \frac{\Theta(\beta - \beta_0) \delta(t - \beta + \beta_0)}
{\left(4 \pi \int_{\beta_0}^\beta{D(\beta')\,{\mathrm d}\beta'}\right)^{3/2}} 
\exp \left(\frac{-(\vec{r} - \vec{r}_0)^2}{4 \int_{\beta_0}^\beta{D(\beta')\,{\mathrm d}\beta'}} \right).
\label{eq:G_sol_beta}
\end{equation}
The Green's function becomes 
\begin{equation}
G(\vec{r},p,t\,|\,\vec{r}_0,p_0) = \frac{\Theta(p_0 - p) \delta\left(t + \int_{p_0}^p{b(p')^{-1}\,{\mathrm d}p'}\right)}
{b(p) \cdot \left(4 \pi \int_{p}^{p_0}{D(p')/b(p')\,{\mathrm d}p'}\right)^{3/2}} 
\exp \left(\frac{-(\vec{r} - \vec{r}_0)^2}{4 \int_{p}^{p_0}{D(p')/b(p')\,{\mathrm d}p'}} \right).
\label{eq:G_sol_app}
\end{equation}
The general solution $n_{\mathrm{p}}(\vec{r},p,t)$ of the transport equation (\ref{transp_eq_large}), 
with a momentum-loss rate given by Eq.\ (\ref{eq:loss_rate}) and for an arbitrary source function $Q$, can be obtained by 
convolving the fundamental solution $G(\vec{r},p,t\,|\,\vec{r}_0,p_0)$ with a source term $Q(\vec{r}_0,p_0,t_0)$, 
\begin{equation}
n_{\mathrm{p}}(\vec{r},p,t) = \iiint{G(\vec{r},p,t\,|\,\vec{r}_0,p_0)Q(\vec{r}_0,p_0,t_0)\,{\mathrm d}t_0\,{\mathrm d}^3r_0\,{\mathrm d}p_0}.
\label{convol_G_Q_app}
\end{equation}
This differential CR proton number density can also be applied to many other astrophysical situations with scalar, momentum-dependent 
diffusion and any type of momentum losses, such as stellar winds, CR diffusion in the interstellar medium, or gamma-ray bursts.

%================================================
\section*{Appendix B: Specific source function for SNR-MC systems\label{app_B}}
%================================================
The source function $Q(\vec{r}_0,p_0,t_0)$ is modeled for four specific SNRs associated with MCs showing gamma-ray emission 
for which data samples from spectral measurements in the X-ray energy range exist. Here, the source spectrum is assumed to be 
of the specific form 
\begin{align}
 Q(\vec{r}_0,p_0,t_0) &= Q_{\mathrm{norm}}\cdot Q_{p}(p_0)\cdot\big[\Theta\left(x_0+l_{\mathrm{c}}/2\right)-\Theta\left(x_0-l_{\mathrm{c}}/2\right)\big] %\nonumber \\
 \cdot\big[\Theta\left(y_0+l_{\mathrm{c}}/2\right)-\Theta\left(y_0-l_{\mathrm{c}}/2\right)\big] \nonumber \\
 &\cdot\big[\Theta(z_0+l_{\mathrm{c}})-\Theta(z_0)\big]\cdot\big[\Theta(t_0)-\Theta(t_0-1)\big],
\label{source_app}
\end{align}
where $Q_{\mathrm{norm}}$ denotes a normalization constant, $Q_{p}(p_0)$ is the spectral shape of the 
low-energy CR protons in terms of the particle momentum $p_0$, and $l_{\mathrm{c}}$ characterizes the 
extent of the emission region. A source function of this type describes emission that is constant over a period of time 
(normalized to the unit time interval of one second) from a cubic emission volume, seen by an observer located 
at the center of a face of the emission volume that coincides with the cloud surface, with a coordinate system 
such that the positive Cartesian $z$-axis is normal to this face and points into the cloud. 
The cubic geometry is chosen over the more physical, spherical geometry for numerical feasibility. 
The volume of the cube-shaped emission region, $l_{\mathrm{c}}^3$, is adapted to the spherical emission volume 
used in the modeling process of the gamma rays in Sect.\ \ref{CR_ion_prof}. 
For the specific source function (\ref{source_app}), the integrations with respect to $\vec{r}_0$, $p_0$, and 
$t_0$, that have to be performed in order to determine the differential CR proton number density (\ref{convol_G_Q_app}), 
\begin{align}
n_{\mathrm{p}}(\vec{r},p,t) &= Q_{\mathrm{norm}} \int_{0}^{\infty} \int_{\mathbb{R}^3} \int_{-\infty}^{\infty} 
\frac{\Theta(p_0 - p)\cdot Q_{p}(p_0)\cdot\delta \left(t + \int_{p_0}^p{b(p')^{-1}\,{\mathrm d}p'} \right)}
{b(p) \cdot \left(4 \pi \int_{p}^{p_0}{D(p')/b(p')\,{\mathrm d}p'}\right)^{3/2}} %\nonumber \\ 
\cdot \exp \left(\frac{-(\vec{r} - \vec{r}_0)^2}{4 \int_{p}^{p_0}{D(p')/b(p')\,{\mathrm d}p'}} \right) \nonumber \\
&\cdot\big[\Theta\left(x_0+l_{\mathrm{c}}/2\right)-\Theta\left(x_0-l_{\mathrm{c}}/2\right)\big] %\nonumber \\
 \cdot\big[\Theta\left(y_0+l_{\mathrm{c}}/2\right)-\Theta\left(y_0-l_{\mathrm{c}}/2\right)\big]\cdot\big[\Theta(z_0+l_{\mathrm{c}})-\Theta(z_0)\big] \nonumber \\
&\cdot\big[\Theta(t_0)-\Theta(t_0-1)\big]\,{\mathrm d}t_0 \, {\mathrm d}^3r_0\,{\mathrm d}p_0,
\label{convol_G_model}
\end{align}
can be done separately. The specific expressions for the actual spectral shapes $Q_{p}(p_0)$ and the normalization 
constants $Q_{\mathrm{norm}}$ used for the astrophysical objects of interest are of no relevance for these integrations. 
Since the Green's function $G(\vec{r},p,t\,|\,\vec{r}_0,p_0)$ is independent of $t_0$, only the time-dependent factor 
of the source function, $Q_{\mathrm{time}}(t_0)=\Theta(t_0)-\Theta(t_0-1)$ has to be integrated with respect to 
$t_0$, yielding 
\begin{equation}
 I_{\mathrm{time}} = \int_{-\infty}^{\infty}{\big[\Theta(t_0)-\Theta(t_0-1)\big]\,\mathrm{d}t_0} = 1. \label{I_time}
\end{equation}
The spatial integration 
\begin{align}
I_{\mathrm{space}} &= \int_{\mathbb{R}^3}{\exp \left(\frac{-(\vec{r} - \vec{r}_0)^2}{4 \int_{p}^{p_0}{D(p')/b(p')\,{\mathrm d}p'}} 
\right)\cdot\big[\Theta\left(x_0+l_{\mathrm{c}}/2\right)-\Theta\left(x_0-l_{\mathrm{c}}/2\right)\big]} 
\cdot \big[\Theta\left(y_0+l_{\mathrm{c}}/2\right)-\Theta\left(y_0-l_{\mathrm{c}}/2\right)\big] \nonumber \\
&\cdot \big[\Theta(z_0+l_{\mathrm{c}})-\Theta(z_0)\big] \,{\mathrm d}^3r_0
\end{align}
results in a product of error functions 
\begin{align}
I_{\mathrm{space}} = \frac{1}{8}\left(4 \pi \int_{p}^{p_0}{D(p')/b(p')\,{\mathrm d}p'}\right)^{3/2} %\nonumber \\
 \cdot \prod_{h=x,y,z+l_{\mathrm{c}}/2}\left[\sum_{j=0}^{1} \mathrm{erf}\left(\frac{l_{\mathrm{c}}/2+(-1)^j\cdot h}
 {\sqrt{4 \int_{p}^{p_0}{D(p')/b(p')\,{\mathrm d}p'}}}\right) \right].
\label{eq:sol_int_space}
\end{align}
Then, one finds the following momentum integral 
\begin{align}
I_{\mathrm{mom}} = \int_0^\infty \frac{Q_{p}(p_0)}{b(p)} \cdot \Theta(p_0 - p) \delta 
\left(t + \int_{p_0}^p{b(p')^{-1}\,{\mathrm d}p'}\right) %\nonumber \\ 
\cdot \prod_{h=x,y,z+l_{\mathrm{c}}/2}\left[\sum_{j=0}^{1} \mathrm{erf}\left(\frac{l_{\mathrm{c}}/2+(-1)^j\cdot h}
 {\sqrt{4 \int_{p}^{p_0}{D(p')/b(p')\,{\mathrm d}p'}}}\right) \right]\,\mathrm{d}p_0.
\label{eq:int_momentum}
\end{align}
In order to solve this integral, first, one has to explicitly evaluate the integrals in the arguments 
of the Dirac distribution and the error functions, respectively. Starting with the integral 
$\int_{p}^{p_0}{D(p')/b(p')\,{\mathrm d}p'}$, keeping the solution as general as possible, the momentum 
dependence of the diffusion coefficient is assumed to be
\begin{equation}
D(p) = D_0 \left(\frac{p}{m_{\mathrm p}c} \right)^k
\label{eq:diff_coeff_def}
\end{equation}
with $D_0 = {\mathrm{const.}}$ and values $1/3 \le k \le 4/3$. Using Eq.\ (\ref{eq:loss_rate}) 
and the dimensionless quantity $\mathfrak{p}\equiv p/(m_{\mathrm{p}}c)$, one obtains 
\begin{equation}
\int_{p}^{p_0}{\frac{D(p')}{b(p')}\,{\mathrm d}p'} 
= \frac{D_0 \, m_{\mathrm p}c}{a_{\mathrm{cc}}} \, \int_{\mathfrak{p}}^{\mathfrak{p}_0} 
{\frac{\left(\mathfrak{p}' \right)^{2+k}}
{a \cdot \left(\mathfrak{p}' \right)^3 + 1}\,{\mathrm d}\mathfrak{p}'},
\label{eq:aux_int_diml}
\end{equation}
where $a = a_{\mathrm{ad}} / a_{\mathrm{cc}}$.
An analytic solution of this integral can be found in \cite{gradshteyn1965} [formula 3.914 number 5], yielding
\begin{equation}
\int_{p}^{p_0}{\frac{D(p')}{b(p')}\,{\mathrm d}p'}
= \frac{D_0 \, m_{\mathrm p}c}{a_{\mathrm{cc}}} \, 
\left[\frac{\left(\mathfrak{p}' \right)^{3+k} \, _{2}F_{1}\left(1, 1+\frac{k}{3}; 2 + \frac{k}{3}; -a 
\cdot \left(\mathfrak{p}' \right)^3 \right)}{3+k} \right]_{\mathfrak{p}' = \mathfrak{p}}^{\mathfrak{p}_0}.
\end{equation}
Here, 
$$_{2}F_{1}(\mathfrak{a},\mathfrak{b};\mathfrak{c};\mathfrak{z})\equiv\frac{\Gamma(\mathfrak{c})}
{\Gamma(\mathfrak{b})\Gamma(\mathfrak{c}-\mathfrak{b})}\int_{0}^{1}{\frac{t^{\mathfrak{b}-1}(1-t)^{\mathfrak{c}-\mathfrak{b}-1}}
{(1-t\,\mathfrak{z})^\mathfrak{a}}\,\mathrm{d}t}$$
denotes the hypergeometric function and 
$$\Gamma(\mathfrak{c})\equiv\int_0^{\infty}{t^{\mathfrak{c}-1}\exp{(-t)}\,\mathrm{d}t}$$ 
the complete Gamma function. 
Substituting this and 
\begin{equation}
 \int_{p_0}^p{b(p')^{-1}\,{\mathrm d}p'} = 
 -\frac{m_{\mathrm p}c}{3\,a_{\mathrm{ad}}}\ln\left(\frac{a_{\mathrm{cc}}+a_{\mathrm{ad}}\mathfrak{p}_0^3}
 {a_{\mathrm{cc}}+a_{\mathrm{ad}}\mathfrak{p}^3} \right)
\end{equation}
into Eq.\ (\ref{eq:int_momentum}), one finds 
\begin{align}
I_{\mathrm{mom}} &= \frac{m_{\mathrm p}c}{ b(\mathfrak{p})} \bigintss_{\mathfrak{p}}^\infty \delta 
\Bigg(t - \frac{m_{\mathrm p}c}{3\,a_{\mathrm{ad}}}\ln\left(\frac{a_{\mathrm{cc}}+a_{\mathrm{ad}}\mathfrak{p}_0^3}
{a_{\mathrm{cc}}+a_{\mathrm{ad}}\mathfrak{p}^3} \right)\Bigg)\cdot Q_{p}(\mathfrak{p}_0) \\ 
&\cdot \prod_{h=x,y,z+l_{\mathrm{c}}/2}\left[\sum_{j=0}^{1}\mathrm{erf}\left(\frac{\left(l_{\mathrm{c}}/2 + (-1)^j\cdot h\right)\sqrt{a_{\mathrm{cc}}(3+k)}}
{\sqrt{4 D_0 \, m_{\mathrm p}c %\, 
\left[\left(\mathfrak{p}' \right)^{3+k} \, _{2}F_{1}\left(1, 1+\frac{k}{3}; 2 + \frac{k}{3}; -a 
\cdot \left(\mathfrak{p}' \right)^3 \right)\right]_{\mathfrak{p}' = \mathfrak{p}}^{\mathfrak{p}_0}}} \right) \right]
\, {\mathrm d}\mathfrak{p}_0. \nonumber 
\label{eq:int_mom_aux}
\end{align}
The zeros of the argument of the Dirac distribution, as a function of the momentum $\mathfrak{p}_0$, are given by 
\begin{equation}
 \mathfrak{p}_0^{\mathrm{zero}}(\mathfrak{p},t) = \Bigg(\mathfrak{p}^3\cdot\exp\left(\frac{3a_{\mathrm{ad}}t}{m_{\mathrm p}c}\right) 
 + \frac{a_{\mathrm{cc}}}{a_{\mathrm{ad}}}\left(\exp\left(\frac{3a_{\mathrm{ad}}t}{m_{\mathrm p}c}\right)-1\right)\Bigg)^{1/3}.
\label{dirac_p}
\end{equation}
Hence, the Dirac distribution can be written as 
\begin{eqnarray}
 \delta \Bigg(t - \frac{m_{\mathrm p}c}{3\,a_{\mathrm{ad}}}\ln\left(\frac{a_{\mathrm{cc}}+a_{\mathrm{ad}}\mathfrak{p}_0^3}
 {a_{\mathrm{cc}}+a_{\mathrm{ad}}\mathfrak{p}^3} \right)\Bigg) %\nonumber \\
 = \frac{\left(a_{\mathrm{cc}}+a_{\mathrm{ad}} \big(\mathfrak{p}_0^{\mathrm{zero}}(\mathfrak{p},t)\big)^3\right)}
 {m_{\mathrm p}c\cdot\big(\mathfrak{p}_0^{\mathrm{zero}}(\mathfrak{p},t)\big)^2} 
 \cdot \delta \big(\mathfrak{p}_0 -\mathfrak{p}_0^{\mathrm{zero}}(\mathfrak{p},t)\big).
\end{eqnarray}
Subsequently, the momentum integral becomes 
\begin{align}
 I_{\mathrm{mom}} &= \frac{a_{\mathrm{cc}}+a_{\mathrm{ad}} \big(\mathfrak{p}_0^{\mathrm{zero}}(\mathfrak{p},t)\big)^3}
 {b\big(\mathfrak{p}\big)\cdot \big(\mathfrak{p}_0^{\mathrm{zero}}(\mathfrak{p},t)\big)^2}
 \cdot Q_{p}(\mathfrak{p}_0^{\mathrm{zero}}(\mathfrak{p},t)) \label{I_mom} \\
 &\cdot \prod_{h=x,y,z+l_{\mathrm{c}}/2}\left[\sum_{j=0}^{1}\mathrm{erf}\left(\frac{\left(l_{\mathrm{c}}/2 + (-1)^j\cdot h\right)\sqrt{a_{\mathrm{cc}}(3+k)}}
 {\sqrt{4 D_0 \, m_{\mathrm p}c %\, 
 \left[\left(\mathfrak{p}' \right)^{3+k} \, _{2}F_{1}\left(1, 1+\frac{k}{3}; 2 + \frac{k}{3}; -a 
 \cdot \left(\mathfrak{p}' \right)^3 \right)\right]_{\mathfrak{p}' = \mathfrak{p}}^{\mathfrak{p}_0^{\mathrm{zero}}(\mathfrak{p},t)}}} \right) \right]. \nonumber
\end{align}
Note that $\Theta \big(\mathfrak{p}_0^{\mathrm{zero}}(\mathfrak{p},t)-\mathfrak{p}\big) = 1$ for all values 
of $\mathfrak{p}_0^{\mathrm{zero}}$ and $\mathfrak{p}$. Then, combining of (\ref{I_time}), (\ref{eq:sol_int_space}) 
and (\ref{I_mom}) leads to the differential CR proton number density of 
a cubic emission source region with edge length $l_{\mathrm{c}}$ for all positions $\vec{r}$ inside the MC, 
with $z\ge0$, at any time $t\ge0$ and for all particle momenta $\mathfrak{p}\in[0.15,\, 0.86] \le \mathfrak{p}_0$ 
\begin{align} \label{eq:sol_transp_eq_app}
n_{\mathrm{p}}(\vec{r},p,t) &= \frac{Q_{\mathrm{norm}}\cdot\left(a_{\mathrm{cc}}+a_{\mathrm{ad}}
 \big(\mathfrak{p}_0^{\mathrm{zero}}(\mathfrak{p},t)\big)^3\right)\cdot Q_{p}\big(\mathfrak{p}_0^{\mathrm{zero}}(\mathfrak{p},t)\big)}
 {8\cdot b(\mathfrak{p})  \cdot \big(\mathfrak{p}_0^{\mathrm{zero}}(\mathfrak{p},t)\big)^2} \\
&\cdot \prod_{h=x,y,z+l_{\mathrm{c}}/2}\left[\sum_{j=0}^{1}\mathrm{erf}\left(\frac{\left(l_{\mathrm{c}}/2 + (-1)^j\cdot h\right)\sqrt{a_{\mathrm{cc}}(3+k)}}
 {\sqrt{4 D_0 \, m_{\mathrm p}c %\, 
 \left[\left(\mathfrak{p}' \right)^{3+k} \, _{2}F_{1}\left(1, 1+\frac{k}{3}; 2 + \frac{k}{3}; -a 
 \cdot \left(\mathfrak{p}' \right)^3 \right)\right]_{\mathfrak{p}' = \mathfrak{p}}^{\mathfrak{p}_0^{\mathrm{zero}}(\mathfrak{p},t)}}} \right) \right]. \nonumber
\end{align}

\end{document}